\documentclass[aps, pre, a4paper, floatfix, twocolumn, twoside, longbibliography, nofootinbib,  amssymb]{revtex4-2}

\usepackage[colorlinks=true, hypertexnames=false]{hyperref}
\PassOptionsToPackage{linktocpage}{hyperref} 
\hypersetup{
  colorlinks  = true, 
  urlcolor = magenta!80!black,
  linkcolor  = blue!60!black, 
  citecolor  = blue!60 
}

\usepackage[T1]{fontenc}
\usepackage[utf8]{inputenc}
\usepackage{epsfig}
\usepackage[english]{babel}
\usepackage[table]{xcolor}
\usepackage{graphicx}
\usepackage{amsmath}
\usepackage{mathtools}
\usepackage{relsize}
\usepackage[percent]{overpic}
\usepackage{bm}
\usepackage{hyperref}
\usepackage{siunitx}
\usepackage{accents}
\usepackage{mathtools}
\usepackage{grffile}
\usepackage[normalem]{ulem}
\usepackage{pgf}
\usepackage{tikz}
\usepackage{times}
\usepackage{subfigure}
\usepackage{nccmath}

\usepackage[capitalize]{cleveref}

\usepackage{dsfont}  

\usepackage{todonotes}

\newcommand{\avg}[1]{{\left<#1\right>}}
\newcommand{\floor}[1]{{\lfloor #1\rfloor}}

\newcommand{\ee}{\mathrm{e}}

\usepackage{mathtools}
\def\multiset#1#2{\ensuremath{\left(\kern-.3em\left(\genfrac{}{}{0pt}{}{#1}{#2}\right)\kern-.3em\right)}}

\usepackage{amsmath}

\newcommand{\g}{\bm{g}}
\newcommand{\f}{\bm{f}}

\newcommand{\cc}{\bm{c}}

\newcommand{\W}{\bm{W}}

\newcommand{\D}{\bm{D}}
\newcommand{\bl}{\bm{\lambda}}

\newcommand{\ER}{Erd\H{o}s--R\'enyi }

\renewcommand{\eqref}[1]{Eq.~\ref{#1}}

\newcolumntype{P}[1]{>{\centering\arraybackslash}p{#1}}
\usepackage{algpseudocodex}
\usepackage{algorithm}

\algnewcommand\Input{\State\textbf{Input:} }
\algrenewcommand\Output{\State\textbf{Output:} }
\algnewcommand\Continue{\State\textbf{continue}}

\usepackage[svgnames]{xcolor}

\begin{document}

\title{Graphs are maximally expressive for higher-order interactions}

\author{Tiago \surname{P. Peixoto}}
\email{peixoto@c3s.uni-frankfurt.de}
\affiliation{Center for Critical Computational Studies, Goethe University Frankfurt, Frankfurt am Main, Germany }
\affiliation{Institute of Computer Science, Goethe University Frankfurt, Frankfurt am Main, Germany }
\affiliation{IT:U, Linz, Austria}
\affiliation{Department of Network and Data Science, Central European University, Vienna, Austria}
\affiliation{Complexity Science Hub, Vienna, Austria}

\author{Leto Peel}
\affiliation{Department of Data Analytics and Digitalisation, Maastricht University, Maastricht, The Netherlands}

\author{Thilo Gross}
\affiliation{Helmholtz Institute for Functional Marine Biodiversity (HIFMB), Oldenburg, Germany}
\affiliation{Alfred-Wegener Institute, Bremerhaven, Germany}
\affiliation{Carl-von-Ossietzky University, ICBM, Oldenburg, Germany}

\author{Manlio De Domenico}
\affiliation{Department of Physics and Astronomy “Galileo Galilei,” University of Padua, Padua, Italy}
\affiliation{Istituto Nazionale di Fisica Nucleare, Sez. Padova, Padua, Italy}
\affiliation{Padua Center for Network Medicine, University of Padua, Padua, Italy}
\affiliation{Padova Neuroscience Center, University of Padua, Padua, Italy}

\begin{abstract}
  We demonstrate that graph-based models are fully capable of representing
  higher-order interactions, and have a long history of being used for
  precisely this purpose. This stands in contrast to a common claim in the
  recent literature on ``higher-order networks'' that graph-based
  representations are fundamentally limited to ``pairwise'' interactions,
  requiring hypergraph formulations to capture richer dependencies. We clarify
  this issue by emphasizing two frequently overlooked facts.
  First, graph-based models are not restricted to pairwise interactions, as they
  easily accommodate interactions that depend simultaneously on multiple
  adjacent nodes. Second, hypergraph formulations are strict special cases of
  more general graph-based representations, as they impose additional
  constraints on the allowable interactions between adjacent elements rather
  than expanding the space of possibilities. We show that key phenomenology
  commonly attributed to hypergraphs---such as abrupt transitions---can, in
  general, be recovered exactly using graph models, even locally tree-like ones,
  and thus do not constitute a class of phenomena that is inherently contingent
  on hypergraphs models. Finally, we argue that the broad relevance of
  hypergraphs for applications that is sometimes claimed in the literature is
  not supported by evidence. Instead it is likely grounded in misconceptions
  that network models cannot accommodate multibody interactions or that certain
  phenomena can only be captured with hypergraphs. We argue that clearly
  distinguishing between multivariate interactions, parameterized by graphs, and
  the functions that define them enables a more unified and flexible foundation
  for modeling interacting systems.
\end{abstract}

\maketitle

\smaller[.5]
\tableofcontents
\newpage

\section{Introduction}
In recent years there has been a surge of interest in modeling interacting
systems via so-called ``higher-order networks'' (HONs), characterized by
hypergraph parameterizations of interactions involving more than two
elements~\cite{battiston_networks_2020,bianconi_higher-order_2021,torres_why_2021,battiston_physics_2021,majhi_dynamics_2022,bick_what_2023,battiston_higher-order_2025,abiad_hypergraphs_2026}.
This literature proposes hypergraphs as the foundation of a general theory of
complex systems, claiming a level of explanatory power unattainable by
graph-based formulations. On this basis, some authors advocate that graphs
should be universally supplanted by hypergraphs as the most elementary
representational
object~\cite{battiston_networks_2020,battiston_higher-order_2025,abiad_hypergraphs_2026},
relegating graph-based models to special cases. The argument that hypergraphs
are essential for modeling multibody interactions rests on the following claims:
\begin{enumerate}
  \item Graphs encode only ``pairwise interactions.''
  \item Hypergraphs encode ``group interactions,'' indivisible interaction units
        with more than two elements that cannot be represented by graphs.
  \item Many systems are better modeled with ``group interactions,'' and hence hypergraphs.
  \item ``Group interactions'' give rise to new phenomenology, not explainable by
        graph-based models.
\end{enumerate}
These claims follow a different line of reasoning and do not engage
substantially with the long tradition in statistical physics of employing
hypergraphs as bipartite factor
graphs~\cite{mezard_information_2009,krzakala_statistical_2015}---a framework
central to the theory of constraint
satisfaction~\cite{mezard_information_2009,krzakala_gibbs_2007}, error
correction~\cite{kabashima_belief_1998,richardson_design_2001}, spin
glasses~\cite{mezard_bethe_2001}, statistical
inference~\cite{zdeborova_statistical_2016}, community
detection~\cite{decelle_inference_2011}, algorithmic
hardness~\cite{moore_nature_2011}, and non-equilibrium disordered
systems~\cite{lokhov_dynamic_2015}.

The notion of ``pairwise interactions'' in the HON literature conflates the
structure of a graph---where edges connect pairs of nodes---with the functional
form of interactions defined on those edges. This conflation suggests that if a
system is represented by a graph, then the interactions of a node with its
neighbors must decompose into independent or additive pairwise terms. However,
graphs define \emph{neighborhoods}, i.e.\ the set of nodes adjacent to a given
node, not the interactions themselves. The functions defined on these
neighborhoods can be arbitrarily complex and multivariate, depending on all
adjacent nodes simultaneously in nonlinear ways. In this work, we show that
once this conflation is resolved, graph-based formulations are revealed to be
not inferior in expressive power, but in fact more general than
hypergraph-based ones.

Specifically, we demonstrate that a hyperedge implies that a certain interaction
plays out between a fixed set of nodes, in a manner coherently experienced by
every member of the set. As a consequence, interactions across different
hyperedges compose at most additively or in an otherwise simple manner.
Hypergraphs therefore impose structure on the interactions that graph-based
models leave open, making hypergraph models the more restrictive class. It
follows that every phenomenon observable in a hypergraph model must also be
observable in a graph-based model.

We examine prominent phenomenological claims attributed uniquely to hypergraph
models, including abrupt transitions in
synchronization~\cite{skardal_higher_2020}, population
dynamics~\cite{burgio_evolution_2020, alvarez-rodriguez_evolutionary_2021,
  wang_evolutionary_2024, civilini_explosive_2024}, epidemic
spreading~\cite{iacopini_simplicial_2019,matamalas_abrupt_2020}, and equilibrium
spin models~\cite{son_phase_2024,robiglio_higher-order_2025}, and show that
identical behavior can be obtained with graph-based formulations that are
asymptotically locally tree-like. We also show that hypergraph frameworks for
dynamical
systems~\cite{mulas2020coupled,gambuzza_stability_2021,gallo2022synchronization}
can be equivalently formulated using multilayer
networks~\cite{de_domenico_mathematical_2013,kivela_multilayer_2014}, obviating
the need for hypergraphs. Beyond the theoretical arguments, we highlight the
lack of empirical evidence supporting the utility of hypergraph-based models:
graph data almost never contain information beyond neighborhoods, and rarely
encode interaction rules explicitly. Such information is typically latent and
must therefore be inferred statistically. Furthermore, as the size of
interaction groups increases, these interactions are increasingly likely to
exhibit sparse internal structure that is accurately represented by standard
graph-based models. Although hypergraph formulations can reasonably offer
interpretable or parsimonious representations in particular cases, they require
explicit justification, rather than being assumed as a default when multivariate
interactions are considered.

Recent work has shown that certain classes of hypergraph-based models with
multivariate interactions can be reformulated as graph-based models with purely
pairwise
interactions~\cite{neuhauser_multibody_2020,sahasrabuddhe_modelling_2021,llabres_reducibility_2026,meloni_higher-order_2026}.
This line of results addresses a different question from the one considered
here. We do not focus on these specific constructions, nor do we argue that
pairwise interactions are universally sufficient. Instead, we establish that
multivariate interactions parameterized by graphs constitute the most general
class, in a framework that is not tied to hypergraph representations. While some
of the aforementioned
studies~\cite{llabres_reducibility_2026,meloni_higher-order_2026} have
identified inconsistencies within the HON literature, they stop short of
systematically re-examining its foundational assumptions, which is the central
objective of our work.

This paper is organized as follows. In Sec.~\ref{sec:pairwise} we demonstrate
that graphs are not confined to ``pairwise interactions.'' In
Sec.~\ref{sec:general} we show that hypergraph formulations are special cases of
graph-based models, and in Sec.~\ref{sec:multilayer} we show how every specific
hypergraph formulation can also be converted into a multilayer network, but not
vice versa. In Sec.~\ref{sec:phenomena} we turn to phenomena that have been
attributed to hypergraph structure, and show they can be reproduced
identically by locally tree-like graphs. In Sec.~\ref{sec:empirical} we discuss
the lack of empirical evidence for the universal applicability of
hypergraph models, and in Sec.~\ref{sec:reducibility} we discuss general aspects
of the reducibility of multivariate interactions. We conclude in
Sec.~\ref{sec:conclusion} with a final discussion.

\section{Graphs specify the domain of interactions}\label{sec:pairwise}

The HON literature frequently asserts that graphs encode only ``pairwise
interactions,'' but does not precisely define this concept. Instead, edges in a
graph are often implicitly assumed to represent individual instances of isolated
``interactions'' between two nodes, independent of all other edges. Whenever
this constraint does not hold, graph representations allegedly become
inadequate. This interpretation, however, does not conform to how network models
can and have actually been defined---across several decades and in multiple
scientific domains.

As an illustration, consider the foundational example of the K\"{o}nigsberg bridge
problem~\cite{euler_solutio_1741}, where we are asked to find a path that
crosses every bridge in a city exactly once. The bridges in question are neither
a graph nor are they a hypergraph. However there is value in describing them
as a graph: Euler showed that this abstraction allows us to say that
there cannot be a solution if more than two of the nodes are of odd
degree. Note that even in this simple example the conclusion is not drawn from
considering pairs of nodes in isolation, but by considering the degrees which
are sizes of neighborhoods of nodes.

Many modern cities contain bridges with three or more terminals, but this does
not automatically make a network representation inadequate. If we require that
every path between bridge terminals must be used exactly once, each multiway bridge
can be modeled as a set of independent links, so that Euler's results remain valid.

A different situation arises if each multiway bridge may be used only once in
total, regardless of the entry or exit terminal. One way to represent this
constraint is to treat every multiway bridge as an additional node, converting
the system into a bipartite network in which each bridge node must be visited
exactly once. This formulation, while still a graph, is completely equivalent to
a hypergraph formulation of the system.

However, this is not the only option. We can preserve the original network
structure by keeping the terminal-to-terminal links independent while assigning
a color to each link and requiring that each color be used only once. If all
links corresponding to a given multiway bridge share the same color, and
different bridges have distinct colors, this colored-link model yields an
equivalent representation of the problem while maintaining the same form of
“pairwise” network abstraction.

As a further example, consider a physical object that is passed among
individuals, such as a dollar bill. Each time two people meet, the bill may
change hands, and modeling this process directly leads to a diffusion equation
on a network. Now suppose that, in some situations, three people meet
simultaneously. In such a gathering, the current holder of the bill may pass it
to either of the other two participants, but not to both at once.

Does the presence of three-person encounters necessitate a hyperedge? No, one
could easily represent a three-person interaction by three pairwise links, where
the transmission rate along each link is reduced to reflect that the bill can be
passed to only one of the two neighbors during a joint meeting. Optionally we
can consider all links that arise from such three-party interactions as a
separate layer of the network where the diffusion rate is globally reduced for
the whole layer. This approach is appealing because it retains the exact
mathematical formulation of the diffusion process and requires only an
adaptation in the weight of edges.

In the following, we show that the observations made in the above examples can
be formalized and generalized. More specifically, we demonstrate that
interactions defined on graphs are not confined to pairwise ones. Instead of an
isolated interaction, in general, the existence of a single edge $(i,j)$
determines at most whether there is a nonzero \emph{conditional dependence} of
the states of node $j$ on node $i$ when the others are kept fixed (and vice versa
for undirected graphs)---anything more specific can only be assessed in the
context of a particular functional definition. In short, a graph specifies which
nodes can interact, but says nothing about the form those interactions take.

A graph is defined as a set of nodes, $\mathcal{V} = \{1,\dots,N\}$, and a set of
edges consisting of pairs of nodes,
$\mathcal{E} \in \mathcal{V}\times\mathcal{V}$. The adjacency matrix,
\begin{equation}
  W_{ij} =
  \begin{cases} 1 & \text{ if edge $(i,j)$  exists},\\
    0 & \text{otherwise,}
  \end{cases}
\end{equation}
serves as convenient representation of the edge set. If the edge pairs are
ordered, the matrix is in general asymmetric, and it may be desirable to
attribute the edges with real weights, so that $W_{ij}\in \mathbb{R}$. We will
switch between these definitions and other minor variations whenever it is
convenient, while keeping the same notation.

Notably, this definition does not yet involve the notion of ``interaction.''
Before we can invoke this concept, we need first to determine what the nodes
represent. Typically, this is done by attributing to each node $i$ its own state
variable $x_i$, which we will assume to be a scalar for simplicity. 
Interactions between such variables are modeled in
a variety of ways, depending on context and the nature of the system, including,
for example,
\begin{align*}
  \dot x_i(t) &= f_i(\bm x(t)), && \hspace{-4.5em}\text{via ODEs},\\
  P(x_i(t)|\bm x(t-1)) &=f_i(x_i(t), \bm x(t-1)), && \\
              &&&\hspace{-4.5em}\text{via Markov chains,} \\
  x_i &= f_i(\bm x), && \hspace{-4.5em}\text{via coupled equations,} \\
  P(x_i|\bm x\setminus x_i) &= f_i(\bm x), && \hspace{-4.5em}\text{via conditional probabilities,}
\end{align*}
where $\bm x = \{x_1,\dots,x_N\}$ denotes all variables in the system, and the
$f_i$ are functions on each node that specify how it interacts with the other
variables. The equations above represent the most general formulation for each
class of system,\footnote{Since a general system like $\dot x_i(t)=f_i(\bm x(t))$ makes no explicit reference to a graph, it is also valid independently of any graph description, or even when the variables are initially associated with other entities within some modelling context, like, for example, the \emph{edges} of some graph. But as we show below, we can always unambiguously associate any such general system with a graph, as we discuss further in footnote~\ref{truism}.} with the exception of the dynamical versions which still admit
time-dependent functions, dependence on delayed variables, etc., but these
further generalizations are not important for our arguments, so we will leave
them aside.

\begin{figure}[t]
  \includegraphics[width=\columnwidth]{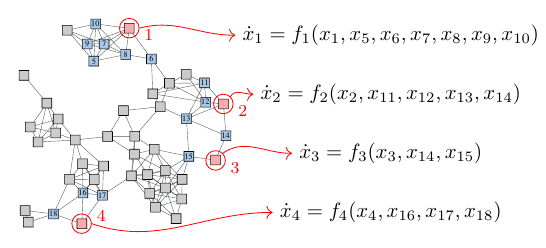}
  \caption{Graphs encode adjacencies (or neighborhoods), which define the domain
    of interactions, not the interactions themselves. The adjacency set
    constrains which variables can influence a node, but only when functions are
    defined on these adjacencies are the interactions specified. Since these
    functions are multivariate in general, they do not need to decompose into
    pairwise terms. This diagram shows a possible instance of the general
    proof-of-concept ODE system of Eq.~\ref{eq:netdyn}, including the equations
    governing the dynamics of the nodes encircled. The adjacent nodes in blue
    (together with the red nodes) define the domain of each
    function.\label{fig:netdyn}}
\end{figure}

A graph-based formulation is useful to further specify the structure of the
interactions without losing any generality, simply by delineating which other
nodes affect a particular node. Crucially, a graph does not encode the
interactions themselves, but only their domain---it constrains which variables
can appear as arguments, but does not define the functional form. For each node
$i$, the graph defines an adjacency set (or neighborhood),
\begin{equation}\label{eq:adj}
  \partial i = \{ j \mid W_{ij} = 1 \},
\end{equation}
and the corresponding set of variables,
\begin{equation}
  \bm x _{\partial i} = \{x_j \mid j \in \partial i\}.
\end{equation}
Based on these adjacency sets, we can further constrain the interactions between
nodes. Taking a system of ODEs as a proof-of-concept (all following arguments
are directly applicable to the other cases above, as well as in equilibrium
systems, as discussed in Appendix~\ref{sec:hamiltonians}), we can in general write
\begin{equation}\label{eq:netdyn}
  \dot x_i = f_i(x_i, \bm x _{\partial i}),
\end{equation}
where we have dropped the explicit time dependence to simplify the notation. We
note that this formulation maintains full expressibility, since we can recover
the general case when we have a complete graph.\footnote{
It is worth emphasizing that Eq.~\ref{eq:netdyn} is neither an assumption nor
a modelling choice, but instead an inherent property of any ODE system given by
$\dot x_i=f_i(\bm x)$. Once we adopt the truism that the edges of a network
should only represent interactions that actually exist, then for any given
$f_i(\bm x)$ we can identify the variables $x_j$ that never directly affect the
value of $x_i$, namely
$u_i = \{j \mid \partial f_i(\bm x)/\partial x_j = 0,\ \forall \bm x\}$, from
which it follows immediately that the neighborhood of a node is composed of
the complementary set, i.e., $\partial i = \{j \mid j \notin u_i\}$, and
hence Eq.~\ref{eq:netdyn} holds. In other words, the adjacency matrix $\W$ is
a property of any conceivable interaction model between variables, which
delineates which interactions matter for the dynamics, not an exogenous,
potentially subjective assumption. Consequently, a network model is not a
distinct class of dynamical systems, but simply one whose interaction
structure is nontrivial and therefore amenable to network analysis.\label{truism}} This shows that a graph constrains but does not
define the interactions between nodes~\cite{peel_statistical_2022}: a function
on a node depends only on the value of the node itself and those of its
neighbors\footnote{One could denote the existence of a self-dependency with a
  self-loop, but we do not take this option, without sacrificing anything
  important.} (see Fig.~\ref{fig:netdyn}).

The only kind of dynamics that can be unambiguously considered ``pairwise'' is
one where node functions decompose into a sum (or other simple combinations) of
independent terms, each involving only two nodes:
\begin{equation}\label{eq:bivariate}
  f_i(x_i,\bm x_{\partial i}) = g_i(x_i) + \sum_{j}W_{ij}h_{ij}(x_i, x_j),
\end{equation}
for some function $h_{ij}$ associated with an edge $(i,j)$, such that the total
contribution to the dynamics is a linear composition of pairwise terms, which do
not require any cooperation between inputs. 

Models with pairwise dynamics of this type only arise in systems that are
inherently linear or in systems that have been strongly simplified by
linearizing around the zero point of the dynamics. If the resulting model is
limited in its phenomenology this is not a result of the underlying graph
framework but rather by the superimposed linearization that can lead to an
oversimplification of the system~\cite{murdoch1969,gross2009}. Consequently,
such pairwise models are the exception rather than the rule.

The historical record easily confirms our argument, since network models have
long employed multivariate functions that do not conform to this pairwise
decomposition.

Perhaps surprisingly, the potential for nonlinear multivariate functions on the
nodes whose order is identical to the node degree was already mentioned in the
very first paper that used the word ``graph'' to denote a network, more than 140
years ago~\cite{sylvester1878}.

More recently, among the earliest graph-based models are the random Boolean
networks introduced by Kauffman in 1969~\cite{kauffman_homeostasis_1969} as a
simplified model of gene regulation. This model describes binary variables
$x_i\in \{0, 1\}$ (i.e.\ a gene is being expressed or not), which evolve
deterministically as
\begin{equation}
  x_i(t) = f_i(\bm x_{\partial i}(t-1)),
\end{equation}
with $f_i$ being a Boolean function defined by a truth table, e.g.
\begin{equation}
\begin{array}{c|c}
  \bm x_{\partial i} & f_i(\bm x_{\partial i}) \\ \hline
  0, 0, \cdots, 0, 0 & 0 \\
  0, 0, \cdots, 0, 1 & 1 \\
  \vdots & \vdots \\
  0, 1, \cdots, 1, 1 & 1 \\
  1, 1, \cdots, 1, 1 & 0,
\end{array}
\end{equation}
with outputs chosen uniformly at random for each input combination and
independently for each node. Many variations on this kind of model have been
investigated over the following
decades~\cite{drossel_random_2008,rozum_boolean_2024}. For example, for the same
class of gene regulation models, an arguably more realistic variation are nested
canalizing functions~\cite{kauffman_random_2003}, defined as
\begin{equation}
  f_i(\bm x_{\partial i}) = \\
  \begin{cases}
    y_1 & \text{ if } x_{\partial i}^{(1)} = z_1 \\
    y_2 & \text{ if } x_{\partial i}^{(1)} \neq z_1 \text{ and  } x_{\partial i}^{(2)} = z_2 \\
    y_3 & \text{ if } x_{\partial i}^{(1)} \neq z_1 \text{ and  } x_{\partial i}^{(2)} \neq z_2  \text{ and  } x_{\partial i}^{(3)} = z_3,\\
    &\vdots\\
    y_k & \text{ if } x_{\partial i}^{(1)} \neq z_1 \text{ and } \dots \text{ and  } x_{\partial i}^{(k)} = z_k,\\
    y_{k+1} & \text{ otherwise, }\\
  \end{cases}
\end{equation}
where $\bm z$ and $\bm y$ are the so-called canalizing values and outputs,
respectively. This function is meant to model how transcription factors bind to
linear portions of the DNA, while being affected by an ordered hierarchy of
inhibitors and promoters. There are also alternative formulations of
multivariate interactions which generalize arbitrary Boolean functions in the
continuous value and time domains, such as those based on Hill
functions~\cite{wittmann_transforming_2009}.

For a broad class of systems, spanning
biological~\cite{li_yeast_2004,wang_process-based_2010} and
social~\cite{watts_simple_2002} contexts, as well as in machine
learning~\cite{rosenblatt_perceptron_1958}, one of the most widely used
multivariate interaction models is the threshold function, given by
\begin{equation}
  f_i(\bm x_{\partial i}) =
  \begin{cases}
    1 & \text{ if } \sum_{j}W_{ij}x_j \ge \tau_i, \\
    0 & \text{ otherwise, }
  \end{cases}
\end{equation}
where $\tau_i$ is a particular threshold value, which can also be expressed as
$f_i(\bm x_{\partial i}) = H( \sum_{j}W_{ij}x_j - \tau_i)$, where
$H(y) = \{1 \text{ if } y \ge 0, \text{ else } 0\}$ is the Heaviside step
function. Besides its relevance for modeling real systems, the threshold
function is so expressive that, if used in a network, it can be used to
represent any Boolean function~\cite{hornik_multilayer_1989}, and if the
Heaviside function is replaced by a sigmoid, it can universally approximate
multivariate continuous
functions~\cite{hornik_multilayer_1989,cybenko_approximation_1989}---a central
fact in the modern theory of machine learning.

None of these graph-based models can be coherently classified as describing
``pairwise interactions,'' since the output of each function depends
collectively and nonlinearly on all inputs, not on any single edge in isolation.
Yet all are defined on ordinary graphs.

Even simple contagion models, such as the susceptible-infected (SI)
dynamics~\cite{pastor-satorras_epidemic_2015}, are not fully pairwise. Consider,
for instance, the case of deterministic infection, defined
for binary values $x_i\in\{0, 1\}$ as
\begin{equation}
  x_i(t+1) = f_i(x_i(t),\bm x_{\partial i}(t)),
\end{equation}
with
\begin{equation}
  f_i(x_i,\bm x_{\partial i}) = x_i + (1-x_i) H\left(\textstyle \sum_{j}W_{ij}x_j - 1\right).
\end{equation}
In this case, knowledge of the state of a single neighbor, if it happens to be
infected, i.e.\ $x_j=1$, is enough to determine the output---so we could say that
the transition $x_i=0 \to 1$ is pairwise. However, no individual input can
guarantee that the output will remain susceptible, i.e.\ $x_i=0 \to 0$, so that
transition requires cooperation between inputs. The situation changes
dramatically when one adds the need for coordination, such as the case of
bootstrap ($k$-core)
contagion~\cite{chalupa_bootstrap_1979,dorogovtsev_k-core_2006,goltsev_k-core_2006},
with
\begin{equation}
  f_i(x_i,\bm x_{\partial i}) = x_i + (1-x_i) H\left(\textstyle \sum_{j}W_{ij}x_j - k\right),
\end{equation}
for some parameter $k>1$. In this case, no single node can determine the
infection state independently from the other ones. Other types of coordination
include different modes of complex
contagion~\cite{centola_complex_2007,centola_spread_2010,ugander_structural_2012,karsai_complex_2014,osullivan_mathematical_2015},
interdependent contagion~\cite{buldyrev_catastrophic_2010}, as well as triadic
percolation~\cite{sun_dynamic_2023}, which exhibits complex dynamical behavior,
including period doubling and route to chaos, and the list can be extended at
will~\cite{gleeson_binary-state_2013}.

Given such varied and well-known examples of multivariate interactions on
graphs, the claim that graphs encode only ``pairwise interactions'' cannot be
sustained.

An alternative---though circular---definition of ``pairwise interaction'' would
be any interaction that can be represented by a graph, with ``higher-order
interactions'' being those that require hypergraphs. This definition, however,
presupposes that these form distinct classes. In the following section, we show
that hypergraphs and graphs support the same class of interactions.

\section{Hypergraphs constrain rather than generalize interactions}\label{sec:general}

The HON literature contrasts graph-based formulations with those based on
\emph{hypergraphs}, a mathematical generalization where edges (or
``hyperedges'') can connect sets of nodes of arbitrary size, not just pairs. As
purely combinatorial objects, hypergraphs generalize graphs: the number of
possible hypergraphs exceeds the number of graphs on the same node set. From
this, one might conclude that hypergraph-based models can express interactions
unavailable to graphs. However, this reasoning conflates combinatorial structure
with functional expressiveness. Since graphs constrain the domain of
interactions rather than defining them, specifying additional structure via
hypergraphs can only impose further constraints, not expand the space of
interactions. In short, hypergraph models are a special case of graph-based
models, not a generalization. We elaborate this point in the following.

Instead of an adjacency matrix, a hypergraph is represented by an adjacency
\emph{tensor},
\begin{equation}
  \lambda_{\cc} =
  \begin{cases}
    1 & \text{ if hyperedge } \cc = (c_1,\dots, c_k) \text{ exists, }\\
    0 & \text{ otherwise, }
  \end{cases}
\end{equation}
where $\cc$ is a set of nodes.

\begin{figure}
  \subfigure[General graph-based formulation]{
    \begin{minipage}{\columnwidth}
    \begin{tabular}[c]{ll}
      \parbox{.33\columnwidth}{\includegraphics[width=.33\columnwidth]{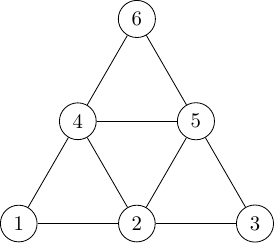}} &
      \parbox{.55\columnwidth}{{\begin{minipage}{.55\columnwidth}
          \vspace{1.2em}
          \begin{fleqn}[1em]
          \[
            \begin{aligned}
              \dot x_1 &= f_1(x_1, x_2, x_4)\\
              \dot x_2 &= f_2(x_1, x_2, x_3, x_4, x_5)\\
              \dot x_3 &= f_3(x_2, x_3, x_5)\\
              \dot x_4 &= f_4(x_1, x_2, x_4, x_5, x_6)\\
              \dot x_5 &= f_5(x_2, x_3, x_4, x_5, x_6)\\
              \dot x_6 &= f_6(x_4, x_5, x_6)
          \end{aligned}
        \]
        \end{fleqn}\\
        \vspace{1em}
        \end{minipage}}}
    \end{tabular}
    \end{minipage}
    }\\
    \subfigure[Hypergraph special case]{
    \begin{minipage}{\columnwidth}
    \begin{tabular}[c]{ll}
      \parbox{.33\columnwidth}{\includegraphics[width=.36\columnwidth]{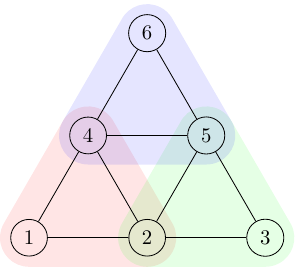}} &
      \parbox{.55\columnwidth}{{\begin{minipage}{.55\columnwidth}
          \vspace{1.5em}
          \begin{fleqn}[1em]
            \[
            \begin{aligned}
              \dot x_1 &= g_1(x_1, x_2, x_4)\\
              \dot x_2 &= g_2(\{x_1, x_2, x_4\},\{x_2, x_3, x_5\})\\
              \dot x_3 &= g_3(x_2, x_3, x_5)\\
              \dot x_4 &= g_4(\{x_1, x_2, x_4\}, \{x_4, x_5, x_6\})\\
              \dot x_5 &= g_5(\{x_2, x_3, x_5\}, \{x_4, x_5, x_6\})\\
              \dot x_6 &= g_6(x_4, x_5, x_6)
            \end{aligned}
            \]
            \end{fleqn}
        \\
        \vspace{1em}
        \end{minipage}}}
    \end{tabular}
    \end{minipage}}
  \caption{(a) Graph-based models are maximally general with respect to a
    particular set of adjacencies. (b) Hypergraph formulations include further
    constraints, amounting to specific choices of the node functions, whose
    structure honors the grouping of adjacent nodes in particular subsets that,
    if combined, form hyperedges (i.e.\ overlapping cliques), as identified by
    shaded regions in (b). A more concrete example is given in
    Fig.~\ref{fig:graph-comp}.\label{fig:hyper-vs-graph}}
\end{figure}

Instead of a set of adjacent neighbors, a hypergraph defines a \emph{set of
  sets} of adjacent neighbors, as one can in principle distinguish the
hyperedges with which they are associated, i.e.
$\Gamma_i = \{ \partial_i(\cc) \mid \lambda_{\cc} =1, i\in\cc\}$, where
$\partial_i(\cc) = \cc \setminus i$ is the set of neighboring nodes
considering only the hyperedge indexed by $\cc$. Adopting this framework, one
may attempt to generalize the proof-of-concept ODE model of Eq.~\ref{eq:netdyn}
to
\begin{equation}\label{eq:hyperdyn}
  \dot x_i = g_i(x_i, \{ \bm{x}_{\cc \setminus i} \mid \lambda_{\cc} =1, i\in\cc\}),
\end{equation}
where $\bm{x}_{\cc \setminus i}$ is the set of variables adjacent to node $i$
via hyperedge $\cc$. In this case, however, we immediately notice that no
generalization has in fact been achieved, since the right hand side of
Eq.~\ref{eq:hyperdyn} must be a multivariate function with a domain given by the
union of all adjacent variables,
$\cup \{ \bm{x}_{\cc \setminus i} \mid \lambda_{\cc} =1, i\in\cc\}$ together with
$x_i$ itself, since this set represents the actual degrees of freedom available
to the dynamics, discounting redundancies. In other words, we can always write
Eq.~\ref{eq:hyperdyn} equivalently as
\begin{align}
  \dot x_i &= f_i(x_i, \cup \{ \bm{x}_{\cc \setminus i} \mid \lambda_{\cc} =1, i\in\cc\}) \\
           &= f_i(x_i, \bm x_{\partial_i}),
\end{align}
where the neighborhood $\partial_i$ in the last equation refers to the projected
adjacency matrix given by
\begin{equation}\label{eq:lproj}
  W_{ij}(\bm \lambda) =
  \begin{cases}
    1 & \text{ if } \sum_{\cc}\bm{1}_{i \in \cc}\bm{1}_{j \in \cc}\lambda_{\cc} > 0,\\
    0 & \text{ otherwise.}
  \end{cases}
\end{equation}
Therefore, any model defined by Eq.~\ref{eq:hyperdyn} maps to the
graph-based model of Eq.~\ref{eq:netdyn}, with the additional constraint that
adjacencies must correspond to projected cliques. While a particular function
$f_i$ may be structured to reflect the adjacency sets defined by the hypergraph
$\bm \lambda$, this is a modeling choice; when it does not hold, the hypergraph
structure provides no additional information.

Moreover, insofar as the structure of the node functions is determined by the
hypergraph,\footnote{Trivial scenarios where the node functions completely
  ignore the hyperedges and are parametrized by the projected graph alone, or
  those where all hyperedges have order two, are nothing but particular
  formulations of graph-based models. In these instances, a hypergraph is not
  operationally distinct from a graph, and contrasting them becomes a
  distinction without a difference.} this imposes additional constraints on the
class of admissible functions. In particular, not only must adjacency sets be
distinguished, but they must also align across adjacent nodes to form
overlapping cliques (see Fig.~\ref{fig:hyper-vs-graph}). By contrast, the
general graph-based formulation relaxes these requirements, both in the
functional form and in how function arguments are connected. Because it does not
enforce an overlapping clique structure, it permits more flexible connectivity
patterns while still representing arbitrarily complex multivariate functions. We
illustrate these differences with concrete examples in the following, and in
Sec.~\ref{sec:mdl} we present an extended analysis based on the enumeration of
model classes.

\begin{figure*}

  \subfigure[Hypergraph-conformant special case]
  {
    \begin{minipage}{.97\columnwidth}
      \centering
      Node adjacencies\\[1em]
      \vbox to 15em {
      \includegraphics[width=\columnwidth]{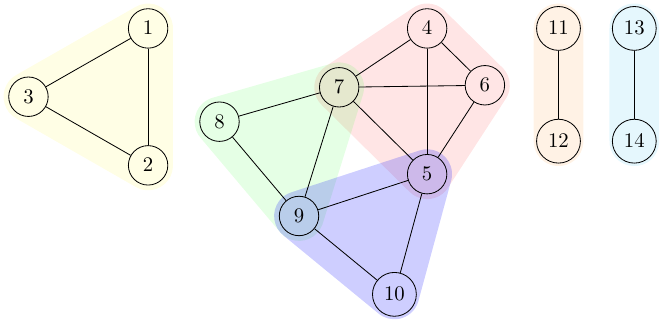}}
      Coupling functions \\
      \vbox to 17em {
       \[
          \begin{aligned}
            \dot x_1 &= h_1(x_1,x_2,x_3) \qquad\hspace{.45em}  \dot x_2 = h_1(x_1,x_2,x_3) \\
            \dot x_3 &= h_1(x_1,x_2,x_3)  \qquad\hspace{.45em} \dot x_4 = h_2(x_4,x_5,x_6, x_7) \\
            \dot x_5 &= h_2(x_4,x_5,x_6,x_7) +  h_3(x_5,x_9,x_{10})  \\
            \dot x_6 &= h_2(x_4,x_5,x_6,x_7) \\
            \dot x_7 &= h_2(x_4,x_5,x_6,x_7) +  h_4(x_7,x_8,x_{9})  \\
            \dot x_8 &= h_4(x_7,x_8,x_{9})  \\
            \dot x_9 &= h_4(x_7,x_8,x_{9}) + h_3(x_5,x_9,x_{10})\\
            \dot x_{10} &= h_3(x_5,x_9,x_{10}) \qquad
            \dot x_{11} = h_5(x_{11},x_{12}) \\
            \dot x_{12} &= h_5(x_{11},x_{12}) \qquad\quad
            \dot x_{13} = h_6(x_{13},x_{14}) \\
            \dot x_{14} &= h_6(x_{13},x_{14}) 
          \end{aligned}
        \]}
        \vspace{.5em}
      \end{minipage}}\hfill   
    \subfigure[Graph-based general case]
    {
      \begin{minipage}{.97\columnwidth}
        \centering
        Node adjacencies\\[1em]
        \vbox to 15em {
          \includegraphics[width=.8\columnwidth]{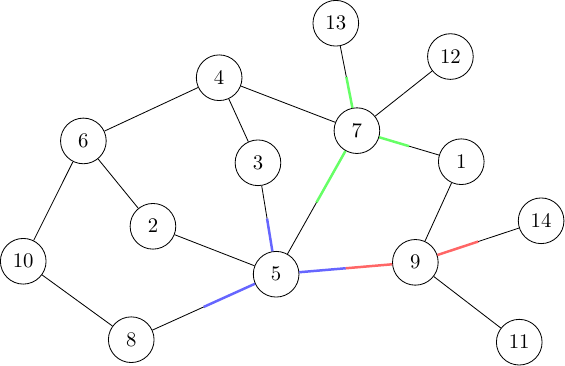}}
        Coupling functions \\
        \vbox to 17em {
        \[
          \begin{aligned}
            \dot x_1 &= h_1(x_1,x_7,x_9) \qquad\hspace{.45em}  \dot x_2 = h_2(x_2,x_5,x_6) \\
            \dot x_3 &= h_3(x_3,x_4,x_5)  \qquad\hspace{.45em} \dot x_4 = h_4(x_3,x_4,x_{6}, x_{7}) \\
            \dot x_5 &= h_5^{(1)}(x_3,x_5,x_8,x_9) +  h_5^{(2)}(x_2,x_5,x_7)  \\
            \dot x_6 &= h_6(x_2,x_4,x_6,x_{10}) \\
            \dot x_7 &= h_7^{(1)}(x_1,x_5,x_7,x_{13}) +  h_7^{(2)}(x_4,x_7,x_{12})  \\
            \dot x_8 &= h_8(x_5,x_8,x_{10})  \\
            \dot x_9 &= h_9^{(1)}(x_5,x_9,x_{14}) + h_9^{(2)}(x_1,x_9,x_{11})\\
            \dot x_{10} &= h_{10}(x_6,x_{8},x_{10}) \qquad
            \dot x_{11} = h_{11}(x_{9},x_{11}) \\
            \dot x_{12} &= h_{12}(x_{7},x_{12}) \qquad\hspace{1.85em}
            \dot x_{13} = h_{13}(x_{7},x_{13}) \\
            \dot x_{14} &= h_{14}(x_{9},x_{14}) 
          \end{aligned}
        \]}
        \vspace{.5em}
      \end{minipage}}

    \caption{Hypergraph parameterizations are special cases of graph-based
      models, and thus offer no generalization. The node adjacencies (top) show
      the skeleton of each model, and the coupling functions (bottom) define the
      interactions. A hypergraph requires \emph{mutual, symmetric membership}:
      if nodes $\{i,j,k\}$ form a hyperedge, a single shared coupling function
      of all three must appear in the equation of every node in that set.
      Panel~(a) satisfies this constraint---the shaded regions mark groups of
      nodes that always appear together in function arguments, forming
      consistent hyperedges (Eq.~\ref{eq:group-wise}). This can also be
      represented by the graph-based model of Eq.~\ref{eq:graph-redux}.
      Panel~(b) uses the same form (Eq.~\ref{eq:graph-redux}), but no hypergraph
      can represent it, since changing the adjacencies and/or the couplings can
      break the mutual membership constraint. For example, $h_5^{(1)}$ groups
      $\{3,5,8,9\}$, but node~3 depends only on $\{3,4,5\}$ via $h_3$, node~8 on
      $\{5,8,10\}$ via $h_8$, and node~9 on $\{5,9,14\}$ via $h_9^{(1)}$---none
      share the same coupling function as node~5, so no hyperedge $\{3,4,8,9\}$
      can exist. We note also that in this particular instance the graph is
      absent of any cliques, which makes it impossible for it to correspond to
      any hypergraph. The colored edge endpoints in~(b) indicate which coupling
      function each edge belongs to, for nodes with more than one. The existence
      of multivariate coupling functions is completely independent from any
      hypergraph structure. Therefore, simultaneously labeling the model of
      panel (a) ``higher-order'' and the one of panel (b) ``pairwise'' or
      ``dyadic'' would be arbitrary.\label{fig:graph-comp}}
\end{figure*}

The objective of most network models is not to remain as general as possible,
but instead to isolate specific mechanisms that constrain the functions $f_i$
being used. For example, for some specific purpose, one may indeed want to
postulate that the functions $f_i$ can be decomposed into a sum of pairwise
terms for each incident edge, i.e.
\begin{equation}\label{eq:graph-wise}
  f_i(x_i,\bm x_{\partial i}) = g_i(x_i) + \sum_jW_{ij}h_{ij}(x_i, x_j).
\end{equation}
And likewise one might consider the analogous decomposition on
hyperedges~\cite{battiston_networks_2020}, i.e.
\begin{equation}\label{eq:group-wise}
  f_i(x_i,\bm x_{\partial i}) = g_i(x_i) + \sum_{\cc}\bm{1}_{i\in \cc}\lambda_{\cc}h_{\cc}(\bm x_{\cc}),
\end{equation}
where each existing group of nodes $\cc$ contributes to a multivariate
interaction term $h_{\cc}$ taking the state of all involved nodes
$\bm{x}_{\cc}$ as arguments. One could then argue that this second class of
models has a larger expressive power than the first (since it is recovered if
all hyperedges are pairs), but it is important not to forget that they are
\emph{both} only special cases of the more general graph-based model of
Eq.~\ref{eq:netdyn}. Furthermore, as we already mentioned, there is a variety of
key examples of widely used graph-based models that do not conform to either
subfamily above, so it is not appropriate to use either of them to make general
statements.

Despite this, one could try to make the case that for each parameterization
based on an adjacency matrix $\W$, one could obtain a more general one by
keeping everything else the same and replacing it by a hypergraph $\bm \lambda$,
as Eq.~\ref{eq:group-wise} does to Eq.~\ref{eq:graph-wise}, to the extent that
makes sense in a particular context (e.g.\ it is not clear how such construction
can be applied to arbitrary Boolean functions, the nested canalizing, or
threshold functions considered previously). However, even when applicable, the
relevance of this argument is quite limited, since we can always escape this
particular parameterization into one that is strictly graph-based and more
general. More specifically, as an example, we can compare the following
graph-based model with the hypergraph-based one of Eq.~\ref{eq:group-wise},
\begin{equation}\label{eq:graph-redux}
  f_i(x_i,\bm x_{\partial i}) = g_i(x_i) + \sum_{l=1}^{m_i}h_i^l(x_i,\bm y_l),
\end{equation}
where to each node $i$ is associated $m_i$ functions $h_i^l$ that act on a
specific subset $\bm y_l$ of the variables adjacent to the same node, so that
$\bm x_{\partial i} = \cup_{l=1}^{m_i} \bm y_l$. If, in Eq.~\ref{eq:group-wise}, we
enumerate the hyperedges incident on node $i$ as $\cc_l (i)$ with $l$ being an
integer, we can make the following mapping
\begin{align}
  W_{ij} &= W_{ij}(\bl), & \text{(via Eq.~\ref{eq:lproj})}\\
  \bm y_l  &= \bm x_{\cc_l(i)\setminus i}, &\\
  h_i^l(x_i,\bm y_l) &= h_{\cc_l(i)}(\bm x_{\cc_l(i)}),& \label{eq:funcs}
\end{align}
to recover Eq.~\ref{eq:group-wise} exactly from Eq.~\ref{eq:graph-redux}.
However, even if we keep the exact same functions of Eq.~\ref{eq:funcs}, up to
the specific choice of arguments, there are typically many other ways to choose
$\W$ and the subsets $\bm y_l$ that are neither expressible by
Eq.~\ref{eq:group-wise}, nor correspond to hypergraph structures, and thus give
rise to completely different systems of interactions.

In Fig.~\ref{fig:graph-comp}a we illustrate an instance of the hypergraph-based
model of Eq.~\ref{eq:group-wise}, showing the coupling functions $h$ that
compose each node function $f_i$, which can also be described according to the
graph-based model of Eq.~\ref{eq:graph-redux}. In Fig.~\ref{fig:graph-comp}b, we
show an alternate instance of the model of Eq.~\ref{eq:graph-redux}, where the
adjacencies and/or the couplings are changed such that it can no longer be
represented by Eq.~\ref{eq:group-wise}.
In fact, even very small variations of Fig.~\ref{fig:graph-comp}a,
such as swapping the contributions of $x_7$ and $x_9$ in the coupling for node
$5$, i.e.
\begin{equation}
  \dot x_5 = h_2(x_4,x_5,x_6,x_9) + h_3(x_5,x_7,x_{10}),
\end{equation}
makes the model no longer expressible via Eq.~\ref{eq:group-wise}, since the two
hyperedges incident on node 5 cease to exist, while it is still describable by
Eq.~\ref{eq:graph-redux}. As with Fig.~\ref{fig:graph-comp}, simultaneously
labeling the hypergraph model of Eq.~\ref{eq:group-wise} ``higher-order'' and
the graph-based one of Eq.~\ref{eq:graph-redux} ``pairwise'' or ``dyadic''
would be arbitrary.

Note that the differences and similarities between both cases in
Fig.~\ref{fig:graph-comp} are not quite visible if we focus only on their
corresponding graph and hypergraph parameterizations, and ignore the functions
that provide the actual couplings between variables---as is often the case when
only the adjacency structure is considered.

In Appendix~\ref{sec:hamiltonians} we describe how the above considerations are
also applicable to equilibrium models described by Hamiltonians.

It is also worth observing that asymmetric interactions are modeled with maximal
generality in graph-based formulations by making edges directed, so that if a
variable $x_j$ appears as an argument of the function for $x_i$, the reciprocal
relationship is not necessarily present. However, the same notion is not as
easily translated for hypergraphs, since this requires making arbitrary and more
restricted decisions about what constitutes an ordering of a group.

The comparisons above reinforce the central point: the most general framework
for network models involves graphs with arbitrary functions defined on
adjacencies. Hypergraph formulations, rather than generalizing this framework,
constitute a constrained subfamily that requires adjacencies to form overlapping
cliques. Without strong empirical evidence that real systems conform to such
constraints, there is no basis for preferring hypergraph formulations over the
more general graph-based approach. We revisit the issue of empirical evidence in
Sec.~\ref{sec:empirical} and model selection in Sec.~\ref{sec:mdl}.

\section{Multilayer networks generalize hypergraphs}\label{sec:multilayer}

The previous section showed that, once coupling functions are considered,
hypergraph formulations do not expand the space of interactions beyond what
graphs already support. In this section we set functions aside and compare the
parametric structures themselves. While a hypergraph cannot in general be
recovered from a single adjacency matrix, we show that multilayer graphs provide
an admissible and strictly more general parametric framework, in which any
hypergraph can be faithfully represented.

Indeed, it is widely recognized---even within the HON
literature~\cite{battiston_networks_2020,bianconi_higher-order_2021,torres_why_2021,battiston_physics_2021,bick_what_2023,battiston_higher-order_2025}---that
any hypergraph can be unambiguously represented as a bipartite factor graph, in
which hyperedges are mapped to factor nodes incident on variable nodes. This
observation is nontrivial, as it allows hypergraphs to be analyzed using
standard graph-theoretic tools. Nevertheless, this possibility is often
downplayed in the HON literature, as discussed in
Sec.~\ref{sec:empirical}.

At the same time, this mathematical equivalence relies on an expansion of the
node set and typically leaves the interpretation of the parameterization
unchanged. It is therefore not the only form of equivalence worth considering.
In the following, we introduce a simple graph-based construction that is not
merely equivalent to hypergraph parameterizations, but strictly generalizes them.

A hypergraph generalizes the structure of a graph, because, from an adjacency matrix
$\W$, one cannot always uniquely recover the hypergraphs that project onto it.
In other words, for a hypergraph defined by the adjacency tensor $\bl$,
we have in general
\begin{equation}\label{eq:proj}
  \lambda_{\cc} \neq \prod_{(i,j) \in \cc \times \cc}W_{ij}.
\end{equation}

\begin{figure}
  \subfigure[]{\includegraphics[width=.32\columnwidth]{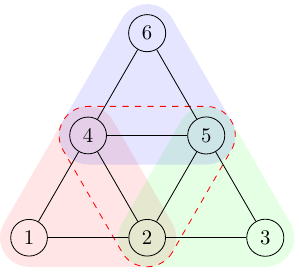}}\hfill
  \subfigure[]{\includegraphics[width=.32\columnwidth]{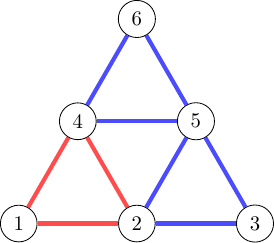}}\hfill
  \subfigure[]{\includegraphics[width=.32\columnwidth]{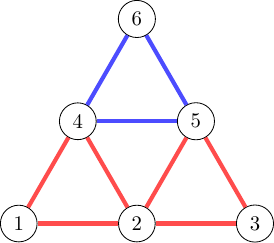}}
  \caption{Hypergraphs define groups of nodes that are not uniquely recovered
    from their graph projections due to the possible existence of cliques that
    are induced by the presence of other cliques. Panel (a) shows an example
    where the existence of hyperedges $(1,2,4)$, $(2,3,5)$, and $(4, 5, 6)$
    means that the projected graph will also contain clique $(2,4,5)$, even
    though that is not included in the hypergraph. However, the bijection is
    fully restored via multilayer graphs, as shown in panels (b) and (c), where
    the requirement is that cliques must contain all edges with the same
    color.\label{fig:cliques}}
\end{figure}

This is true since the existence of some cliques in a graph can induce the
presence of other cliques, as shown in Fig.~{\ref{fig:cliques}}a, and there is
no way to distinguish between the nominal (i.e.\ those directly specified by
$\bl$) and actual cliques (those induced by $\W$). This is the same
information loss encountered in one-mode projections of bipartite graphs, where
shared neighbors induce cliques that cannot be uniquely decomposed. Putting it
differently, an equal sign in Eq.~\ref{eq:proj} would define a system of
equations which is, in general, overdetermined in $\W$, and hence typically
does not have a solution. We see this by noting that for a fixed order $k>2$, the left-hand side of
Eq.~\ref{eq:proj} has ${N\choose k}$ entries, whereas the right-hand side has
only ${N\choose 2}$ free parameters.

However, a rather simple bijection between graphs and hypergraphs can be
recovered by annotating the edges, and separating them into
layers~\cite{de_domenico_mathematical_2013,kivela_multilayer_2014}
$l \in \{1,\dots, L\}$, with $W_{ij}^l$ being the adjacency matrix at layer $l$,
so that
\begin{equation}\label{eq:multilayer}
  \lambda_{\cc} = \sum_{l=1}^{L}\prod_{(i,j) \in \cc \times \cc}W_{ij}^l,
\end{equation}
can always be fulfilled, given a suitable layer decomposition. The most direct
one is when each layer isolates the edges that correspond to each individual
hyperedge, but a decomposition into far fewer layers is often possible (see
Fig.~\ref{fig:cliques}b and Fig.~\ref{fig:cliques}c). The only requirement is
that the cliques that otherwise would have been induced by the nominal ones must
have at least one of their edges in a layer distinct from the rest.\footnote{In the
  case of hypergraphs of mixed order, one would need to add the constraint that
  labelled cliques must be maximal to correspond to a hyperedge, so that their
  subgraphs are not considered.} 

We note that Eq.~\ref{eq:multilayer} should no longer be considered a projection
of a hypergraph into a multilayer graph, since there are typically many different
layer decompositions that map to the same hypergraph. This is particularly true in
the weighted case with $\lambda_{\cc}\in \mathbb{R}$ and
$W_{ij}\in \mathbb{R}$, since with a suitable but fixed layer decomposition the
system will still be underdetermined in $\W$, and hence there will be
infinitely many weighted multilayer networks that map to the same weighted
hypergraph---thus constituting a projection in the opposite direction: from
multilayer graphs to hypergraphs.

The above fact represents important modeling opportunities, since it allows one
to generalize each hyperedge, otherwise representing a monolithic unit, into a
subgraph of arbitrary shape and weight distribution.

As a counterpoint to the above generalization, one could argue that it is
possible to generalize multilayer graphs with multilayer hypergraphs, where the
hyperedges are annotated. However, applying the same argument above means that
this can always be mapped back to multilayer graphs with an expanded layer
decomposition. The result is that there is no structure represented by a
hypergraph, even a multilayer one, that cannot be represented by a single
multilayer graph.

The multilayer generalization of hypergraphs gives us a context-independent way
of converting any hypergraph model into a graph-based one, without changing
anything else in the model other than replacing $\bl$ by $\W$ via
Eq.~\ref{eq:multilayer}. For example, this could be done to
Eq.~\ref{eq:group-wise} to obtain an entirely equivalent dynamical model that
does not rely on hypergraphs. The two formulations are identical, differing only
in whether adjacencies are encoded via a tensor or a multilayer matrix, further
illustrating that the distinction on this basis between ``higher-order'' and
``pairwise'' does not reflect an actual difference in the underlying model.
However, one should not forget that this generalization is not unique, and many
others are possible by adding parameters to the interaction functions or
changing their structure, as we discussed previously.

\section{Phenomenology misattributed to hypergraph modeling}\label{sec:phenomena}

A common claim in the HON literature is that hypergraph models introduce
distinct classes of dynamics that are not accessible to graph-based
formulations~\cite{battiston_networks_2020,bianconi_higher-order_2021,torres_why_2021,battiston_physics_2021,bick_what_2023,battiston_higher-order_2025}.
While the previous sections already show that graph-based models generalize
those constructed via hypergraphs, in this section we examine specific claims
independently. We show that behaviors attributed to hypergraph models can be
reproduced exactly by graph-based models that lack any clique structure, let
alone hyperedges, or by multilayer graphs, without appealing to the
generalization discussed above.

In Sec.~\ref{sec:mean-field} we show that the mean-field calculations used to
justify many of the claims in the HON literature do not distinguish between
hypergraph and graph-based models. In Sec.~\ref{sec:abrupt} we show that the
abrupt transitions reported in synchronization and contagion belong to the same
class of transitions found in graph-based models. In Sec.~\ref{sec:stability} we
show that the stability analysis of ecosystem models is unrelated to hypergraph
structures. Finally, in Sec.~\ref{sec:multilayer_dynamics} we show that a broad
class of hypergraph dynamics can be equivalently formulated as special cases of
multilayer dynamics.

\subsection{Unsuitability of mean-field calculations}\label{sec:mean-field}

A typical form of dynamical models based on hypergraphs studied in the
literature amounts to a special case of Eq.~\ref{eq:group-wise} with
homogeneous functions on all hyperedges, i.e.
\begin{multline}\label{eq:hypercrap}
  \dot x_i = f(x_i) + \sum_j\lambda_{ij}f(x_i,x_j) \\
  + \sum_{jk}\lambda_{ijk}f(x_i,x_j,x_k) \\
  + \sum_{jkl}\lambda_{ijkl}f(x_i,x_j,x_k,x_l) + \cdots,
\end{multline}
where $\lambda_{i,\dots,m}\in\mathbb{R}$ are weighted adjacency tensors of
specific orders, and $f(x_i,\dots,x_m)$ are multivariate functions acting on
each hyperedge $(i,\dots,m)$. (Like before, we will use a system of ODEs as a
proof-of-concept, but our arguments generalize beyond this particular setting.)
In the general case, the tensors are assumed to be sampled from some
distribution $P(\bm \lambda)$, characterized by the ensemble averages,
\begin{equation}
  \avg{\lambda_{i,\dots,m}} = \sum_{\bl} \lambda_{i,\dots,m} P(\bl).
\end{equation}

A typical technique to analyze such systems is to employ a mean-field ansatz,
where the adjacencies in Eq.~\ref{eq:hypercrap} are approximated by their
ensemble averages, $\lambda_{i,\dots,m} \to \avg{\lambda_{i,\dots,m}}$. In the
homogeneous mean-field case, these ensemble averages are assumed to be identical
for entries of the same order, i.e.
\begin{align}
  \avg{\lambda_{i,j}} &= K_1, \\
  \avg{\lambda_{i,j,k}} &= K_2, \\
  \avg{\lambda_{i,j,k,l}} &= K_3,
\end{align}
and so on for higher orders. In the binary case $\lambda_{\cc}\in \{0,1\}$,
this approximation corresponds to the modeling assumption that each potential
hyperedge has the same independent probability of existing---an extension of the
\ER model for hypergraphs, which ignores correlations in the placement of
hyperedges. Based on this approximation, the original dynamics becomes
equivalent to all-to-all couplings of the form:
\begin{multline}\label{eq:mf}
  \dot x_i = f(x_i) + K_1\sum_jf(x_i,x_j) \\
  + K_2\sum_{jk}f(x_i,x_j,x_k) \\
  + K_3\sum_{jkl}f(x_i,x_j,x_k,x_l) + \cdots.
\end{multline}
In this kind of system, every node interacts with every other in the same way,
and this symmetry can often be used to reduce the $N$ coupled differential
equations into a single self-consistency equation for a scalar order parameter
that describes some global average at a fixed point or some other attractor of
interest. For a significant class of systems, this approximation of the order
parameter becomes asymptotically exact in the thermodynamic limit, $N\to\infty$.

This kind of modeling assumption and mean-field calculation have been used to
study a variety of dynamical systems parameterized by hypergraphs, including
synchronization~\cite{skardal_higher_2020,kovalenko_contrarians_2021}, epidemic
spreading~\cite{matamalas_abrupt_2020}, population
dynamics~\cite{burgio_evolution_2020, alvarez-rodriguez_evolutionary_2021,
  wang_evolutionary_2024, civilini_explosive_2024}, models of decision
making~\cite{march-pons_symmetry_2026}, triadic
percolation~\cite{sun_higher-order_2024}, and equilibrium spin
systems~\cite{son_phase_2024,robiglio_higher-order_2025}. The observed behavior
in this setting has been used to claim that hypergraphs give rise to
qualitatively different behavior that is
``invisible''~\cite{battiston_networks_2020,battiston_physics_2021,battiston_higher-order_2025}
to network-based models, typically involving abrupt transitions of the order
parameter.

The key observation that we make is that there are simple alternative model
formulations, not based on hypergraphs, that map to the exact same mean-field of
Eq.~\ref{eq:mf}. For example, we can consider a monolayer graph with adjacency
matrix $\W$, and a dynamics given by
\begin{multline}\label{eq:graph}
  \dot x_i = f(x_i) + \kappa_1\sum_jW_{ij}f(x_i,x_j) \\
  + \kappa_2\sum_{jk}W_{ij}W_{ik}f(x_i,x_j,x_k) \\
  + \kappa_3\sum_{jkl}W_{ij}W_{ik}W_{il}f(x_i,x_j,x_k,x_l) + \cdots,
\end{multline}
where the same multivariate functions $f(x_i,\dots,x_m)$ operate on subsets of
neighbors adjacent to node $i$, without requiring them to form a clique. Once we
approximate the adjacency entries by their ensemble averages,
$W_{ij} \to \avg{W_{ij}} = \rho$, we recover the exact same mean-field equations
of Eq.~\ref{eq:mf}, by setting $\kappa_l\rho^l = K_l$. This means that even
though the original model is supposed to represent structures such as in
Fig.~\ref{fig:graph-comp}a, the mean-field calculation is actually describing
the more general and less constrained model of Fig.~\ref{fig:graph-comp}b, which
is not typically characterized by hyperedges.

In the sparse case with $\rho = O(1/N)$, this homogeneous mean-field represents
a situation where the resulting graph is locally tree-like (as in
Fig.~\ref{fig:graph-comp}b), and hence is absent
of any cliques in the thermodynamic limit. Yet, even though the models of
Eq.~\ref{eq:hypercrap} and Eq.~\ref{eq:graph} have different \emph{microscopic}
dynamics, their \emph{macroscopic} behavior, as far as they are captured by the
mean-field ansatz, are fully identical. Therefore analyses of this kind that are
intended to show that hypergraph models give rise to a particular large scale
behavior actually prove the opposite: that the existence of hyperedges (or
``group interactions'') is not a necessary---or even relevant---ingredient.

\subsection{Abrupt transitions}\label{sec:abrupt}

\begin{figure*}
  \subfigure[Bootstrap ($k$-core) percolation]{\includegraphics[width=.245\textwidth]{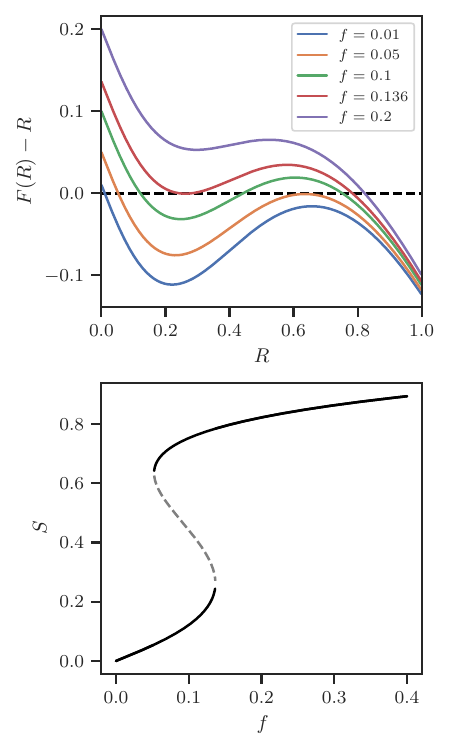}}
  \subfigure[Interdependent percolation]{\includegraphics[width=.245\textwidth]{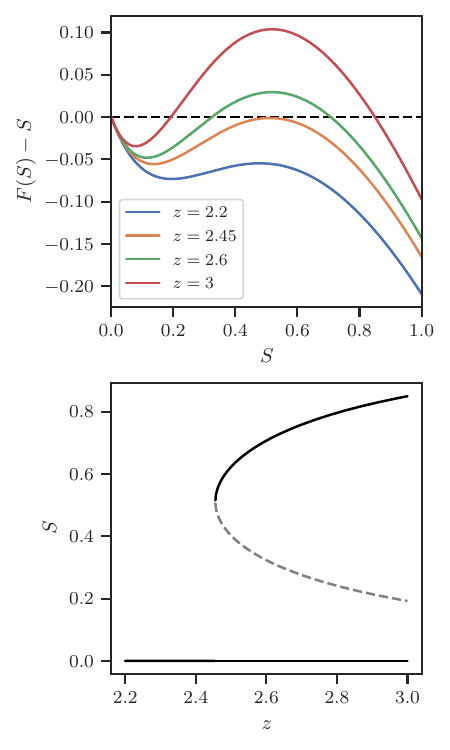}}
  \subfigure[Higher-order synchronization]{\includegraphics[width=.245\textwidth]{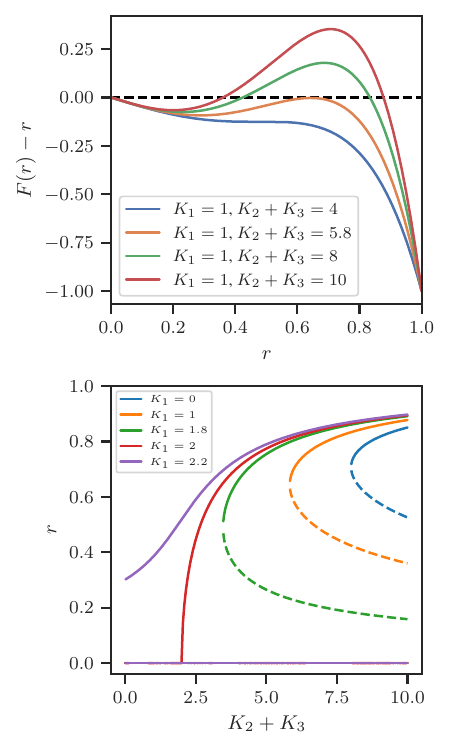}}
  \subfigure[Simplicial contagion]{\includegraphics[width=.245\textwidth]{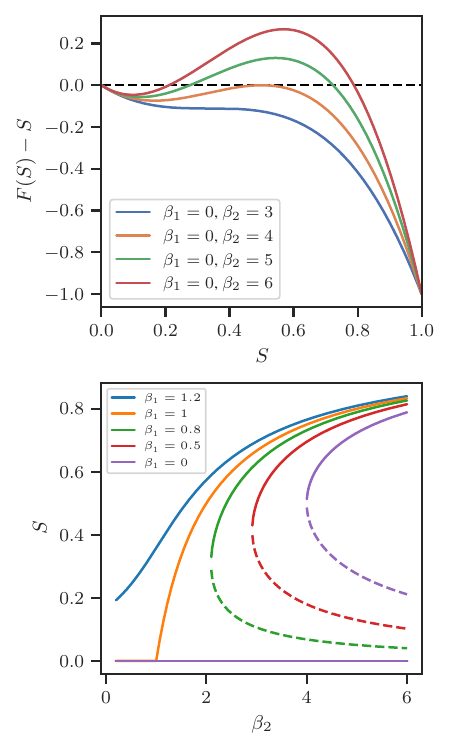}}
  \caption{Abrupt transitions in homogeneous hypergraph models are identical to
    locally tree-like graph models with the same interaction functions, and
    belong to the same phenomenology class of graph-based abrupt transitions.
    The panels show the saddle point bifurcations (top) and corresponding abrupt
    transitions of the order parameter (bottom) for (a) bootstrap ($k$-core)
    percolation for $k=3$ on a \ER (ER) model with mean degree $z=5$, (b)
    interdependent percolation for an ER model and various mean degree values,
    $z$, as indicated in the legend, (c) higher-order synchronization with
    coupling parameters given by the legend, which corresponds to an ER model
    with mean degree $z$, such that $K_l \to K_l (N/z)^l$ in Eq.~\ref{eq:kuramoto}, and (d)
    simplicial contagion, corresponding to an ER model with mean degree $z$,
    such that $\beta_k\to \beta_k(N-1)/[z {N-1\choose k-1}]$ in
    Eq.~\ref{eq:graph_inf}, for $k\in\{1,2\}$, and $\mu=1$. The dashed lines in
    the bottom figures show the unstable fixed points.\label{fig:trans}}
\end{figure*}

The fact that the models of Eq.~\ref{eq:hypercrap} and Eq.~\ref{eq:graph} map to
the same mean-field means that the only relevant ingredient are the multivariate
functions $f$ that take into account the state of multiple neighboring nodes. In
fact, the onset of abrupt transitions has been considered in many models where
these functions reflect the simultaneous occurrence of independent events
related to the set of inputs. Perhaps the simplest and oldest example of this is
given by bootstrap ($k$-core)
percolation~\cite{chalupa_bootstrap_1979,dorogovtsev_k-core_2006,goltsev_k-core_2006},
closely related to the threshold model of Watts~\cite{watts_simple_2002},
defined dynamically on binary-state nodes, $x_i\in \{0, 1\}$, as
\begin{equation}
  x_i(t+1) = x_i(t) + \left[1-x_i(t)\right]H\left(\textstyle\sum_jW_{ij}x_j(t) - k\right),
\end{equation}
for some parameter $k$, where $H(y)$ is the Heaviside step function. For $k=1$,
ordinary percolation (i.e.\ SI dynamics) is recovered, since at least one
neighbor needs to belong to the infected cluster for a node to belong as well.
However, for $k > 1$ cooperative behavior among neighbors is needed for the
onset of the percolating cluster, since at least this many neighbors need to
simultaneously belong to it. The mean-field analysis for the steady state
$t\to\infty$ is done in detail in Refs.~\cite{dorogovtsev_k-core_2006,
  baxter_bootstrap_2010}, and amounts to a self-consistency equation $R=F(R)$
with
\begin{equation}
  F(R)= f + (1-f)\sum_{q=k}^{\infty}\frac{(q+1)P(q+1)}{z}\sum_{l=k}^q{q\choose l}R^l(1-R)^{q-l},
\end{equation}
where $f=\frac{1}{N}\sum_ix_i(0)$ is the initial fraction of active nodes,
$P(q)$ is the degree distribution with mean $z$, and the order parameter is given by
\begin{align}
  S &= \lim_{t\to\infty}\frac{1}{N} \sum_{i}x_i(t) \\
    &= f + (1-f)\sum_{q=k}^{\infty}P(q)\sum_{l=k}^q{q\choose l}R^l(1-R)^{q-l}.
\end{align}
As shown in Fig.~\ref{fig:trans}a, for some parameter values the function $F(R)$
has an inflection which can cause three fixed points to co-exist, only two of
which are stable. As the control parameter $f$ changes, one stable fixed point
undergoes a saddle point bifurcation with the unstable one, leaving the
remaining fixed point as the only stable solution, resulting in a discontinuous,
second-order transition. This happens twice as the dynamics enters and leaves a
bistable coexistence regime.

A different but qualitatively comparable case is interdependent
percolation~\cite{buldyrev_catastrophic_2010}, with a dynamics given by
\begin{equation}
  x_i(t+1) = H\left(\textstyle\sum_jW_{ij}^{(1)}x_j(t)-1\right)  H\left(\textstyle\sum_jW_{ij}^{(2)}x_j(t)-1\right),
\end{equation}
parameterized in the simplest case by a multilayer matrix $W_{ij}^{(l)}$ with two
layers, $l\in\{1,2\}$. The above dynamics means that in order for a node to
belong to the percolating cluster, at least one neighbor in each layer must
belong to the cluster, simultaneously for each layer. If each layer is an \ER
network with the same mean degree $z$, the mean-field calculation leads
to a self-consistency $S=F(S)$ with~\cite{son_percolation_2012}
\begin{equation}
  F(S) = \left(1-e^{-z S}\right)^2,
\end{equation}
which is almost the same equation as standard percolation, except the right-hand
side is squared---reflecting that co-incidence of two independent events that
the dynamics stipulates. Although not identical, the macroscopic behavior is
comparable to bootstrap percolation, since, as can be seen in
Fig.~\ref{fig:trans}b, the function $F(S)$ also possesses an inflection that
leads to saddle point bifurcation and a discontinuous transition.

Both cases above represent mechanisms where abrupt transitions are obtained due
to the requirement of different kinds of coordination among neighbors of a node,
without hypergraphs playing any role. Despite their importance and being very
well known in statistical physics and networks literature, these processes are
almost never referenced in the HON literature. We argue that instances of abrupt
transitions claimed to be due to hypergraph parameterizations are just other
instances of similar phenomena, where the hypergraph parameterization is largely
irrelevant. We consider specific examples in the following.

\subsubsection{Synchronization}\label{sec:synch_multivariate}

Perhaps the clearest example of the conflation described above is the hypergraph
synchronization model of Ref.~\cite{skardal_higher_2020}, an extension of the
Kuramoto model~\cite{arenas_synchronization_2006} with multivariate terms,
defined as
\begin{multline}\label{eq:kuramoto}
  \dot\theta_i = \omega_i + \frac{K_1}{z_1}\sum_j\lambda_{ij}\sin(\theta_j - \theta_i) \\
  + \frac{K_2}{2z_2}\sum_{jk}\lambda_{ijk}\sin(2\theta_j - \theta_k -\theta_i) \\
  + \frac{K_3}{6z_3}\sum_{jkl}\lambda_{ijkl}\sin(\theta_j+\theta_k-\theta_l-\theta_i),
\end{multline}
where $\omega_i$ is the inherent frequency of node $i$, $\lambda_{ij}$,
$\lambda_{ijk}$, are $\lambda_{ijkl}$ are binary adjacency tensors of respective
order, $z_1$, $z_2$, and $z_3$ are the average number of hyperedges incident on
nodes, and $K_1$, $K_2$, and $K_3$ are coupling strengths. For the order
parameter given by $r \ee^{i\phi}$ the authors used a mean-field ansatz to
obtain a self-consistency equation for the steady-state amplitude given by
$r=F(r)$, with
\begin{equation}
  F(r) = \frac{K_1}{2} r(1-r^2) +  \frac{K_2+K_3}{2} r^3(1-r^2).
\end{equation}
Like before, as can be seen in Fig.~\ref{fig:trans}c, whenever $K_2 + K_3>0$ this
function has an inflection with respect to the diagonal, and the system
undergoes the same type of saddle node bifurcation as before as the value of
$K_2 + K_3$ changes, as well as a pitchfork bifurcation when $K_1$ is
varied~\cite{skardal_higher_2020} (not shown).

However, as we have described previously, this behavior is completely unrelated
to the hypergraph parameterization, and remains identical if we replace it with a
tree-like graph in a manner identical to Eq.~\ref{eq:graph}, i.e.\ by replacing
$\lambda_{ijk} = W_{ij}W_{ik}$ and $\lambda_{ijkl} = W_{ij}W_{ik}W_{il}$ in
Eq.~\ref{eq:kuramoto}. For example, if we consider $\W$ to be an \ER network
with mean-degree $z$, we obtain the same mean-field for Eq.~\ref{eq:kuramoto} as
long as we re-scale $K_l \to K_l (N/z)^l$. Therefore, the abrupt transition can
be attributed uniquely to the requirement that multiple neighbors must
synchronize simultaneously for a node to follow them, rather than forming
cliques, let alone hyperedges, or simplices, as originally
claimed~\cite{skardal_higher_2020}.

The defining property of mean-field calculations, especially in the homogeneous
case, is that they do not retain information about correlations between nodes.
If such computations could incorporate correlations simply by adopting a
hypergraph formulation, this limitation would not arise in the first place. It
is therefore unsurprising that the resulting dynamics are equivalent to those on
an uncorrelated random graph. What is striking, however, is that this type of
argument has been presented as evidence for the role of structural correlations
among nodes in the form of hyperedges.

The same kind of mean-field calculation, or equivalently, simulations on
uniformly mixed hypergraphs, have been used to justify the role of these
parameterizations in abrupt transitions and other phenomena, not only in
synchronization, but also population dynamics of
cooperation~\cite{burgio_evolution_2020, alvarez-rodriguez_evolutionary_2021,
  wang_evolutionary_2024, civilini_explosive_2024}, epidemic
spreading~\cite{iacopini_simplicial_2019,matamalas_abrupt_2020,lucas_simplicially_2023,zhao_susceptible-infected-recovered-susceptible_2024},
triadic percolation~\cite{sun_higher-order_2024}, models of decision
making~\cite{march-pons_symmetry_2026}, and equilibrium spin
dynamics~\cite{robiglio_higher-order_2025}. All these cases suffer from the same
problem outlined above: the behavior described can be obtained by tree-like
graphs defined with the same multivariate functions, and the hypergraph
structure plays no role.

Several works also examine the effects of correlations between hyperedges,
typically through numerical simulations or heterogeneous mean-field
calculations~\cite{skardal_higher_2020,landry_effect_2020,gambuzza_stability_2021,larock_encapsulation_2023,skardal_multistability_2023,kim_contagion_2023,zhang_higher-order_2023,burgio_triadic_2024,malizia_hyperedge_2025,malizia_disentangling_2025,malizia_nested_2026}.
However, these studies systematically omit comparisons with appropriately
constructed graph-based representations (such as tree-like or correlated graphs
defined on the same multivariate interaction functions, for example with
neighborhood overlaps), thereby limiting the conclusions that can be drawn
regarding the strict necessity of a hypergraph structure.

We stress that we are not claiming that every dynamics that takes place on a
hypergraph is always macroscopically equivalent to what can be obtained with a
tree-like graph model. Instead, our argument is that: 1. this is true for many
specific claims of abrupt transitions in hypergraph models; 2. the effects of
hypergraph-based correlations should be considered separately from the
existence of multivariate interactions; and 3. claims of unique behaviors
observed with hypergraph models cannot be justified without a direct comparison
with graph-based models with multivariate functions that possess trivial and
nontrivial correlations of their own.

More generally, the presence of structural correlations is expected to influence
a wide range of dynamical processes on graphs, as has long been established in
key
settings~\cite{shirley_impacts_2005,newman_random_2009,osullivan_mathematical_2015},
and this must also hold for correlations parametrized as hypergraphs~\cite{skardal_higher_2020,landry_effect_2020,gambuzza_stability_2021,larock_encapsulation_2023,skardal_multistability_2023,kim_contagion_2023,zhang_higher-order_2023,burgio_triadic_2024,malizia_hyperedge_2025,malizia_disentangling_2025,malizia_nested_2026}. Such
effects, however, should be regarded as secondary relative to the role of
multivariate cooperative interactions among neighbors---an aspect that is
separable from the existence of cliques or hyperedges---and they need to be
contrasted to corresponding correlations under a graph-based formulation (such as
neighborhood overlaps). Nonetheless, the HON literature often claims that the
structures specified by hypergraphs are fundamentally distinct from those
obtained by planting cliques in
graphs~\cite{battiston_networks_2020,battiston_physics_2021,battiston_higher-order_2025},
or even from locally tree-like configurations, and that they give rise to
qualitatively different behavior. At present, however, this claim remains
largely unsubstantiated.

\subsubsection{Contagion}\label{sec:contagion}

A further example of the same issue described above is the simplicial contagion
model of Ref.~\cite{iacopini_simplicial_2019}. In that work, the authors
contrast usual models of complex
contagion~\cite{centola_complex_2007,centola_spread_2010,ugander_structural_2012,karsai_complex_2014,osullivan_mathematical_2015},
where neighbors become infected if the infected neighborhood exceeds a
threshold, with a hypergraph model where a node becomes infected only if all
neighbors belonging to the same hyperedge become infected. More precisely, they
consider a continuous time SIS model defined by a Markov chain,
\begin{multline}
  P(x_i(t')\mid x_i(t),\bm x_{\partial i}(t))
  = \\
  x_i(t)\left[(1-\delta t\, \mu)^{x_i(t')}\left(\delta t\,\mu\right)^{1-x_i(t')}\right] + \\
  \left(1-x_i(t)\right)\left[{\left(\delta t\, \nu_i\right)}^{x_i(t')}{\left(1-\delta t\, \nu_i\right)}^{1-x_i(t')} \right],
\end{multline}
where $t' = t+\delta t$ is the time after an infinitesimal step $\delta t$, $\mu$ is a
uniform recovery rate, and with an overall infection rate given by
\begin{equation}\label{eq:infrate}
  \nu_i = \sum_{\cc}\bm{1}_{i\in\cc}\lambda_{\cc}\beta_{|\cc|}\prod_{j\in \cc}x_j(t),
\end{equation}
where $\beta_k$ is the specific infection rate associated with a hyperedge
$\cc$ of order $|\cc| = k$. In the above dynamics, an infection of a node
only occurs if all its neighbors that belong to the same hyperedge are infected.

The authors of Ref.~\cite{iacopini_simplicial_2019} put an emphasis on the
distinction between the model above and those modeled via graphs---but as we
have shown before in many examples, this is a distinction without a difference,
since the exact same microscopic dynamics represents a special case of a
graph-based model, with the hyperedges representing only a particular
configuration. However, as in the case of synchronization, the phenomenology
described in Ref.~\cite{iacopini_simplicial_2019} is in fact unrelated to any
particular hypergraph structure, as we show below.

In Ref.~\cite{iacopini_simplicial_2019}, the homogeneous mean-field
calculation was obtained for the case of homogeneous mixing by replacing
$\lambda_{\cc}=z_{|\cc|}/{N - 1\choose |\cc| - 1}$ in Eq.~\ref{eq:infrate}, resulting in a
self-consistency equation $S=F(S)$, with
\begin{equation}\label{eq:mf_simplicial}
  F(S) = \frac{(1-\rho)}{\mu}\sum_{k}\beta_kz_kS^k,
\end{equation}
for the order parameter
\begin{equation}
S = \lim_{t\to\infty}\frac{1}{N}\sum_i x_i(t).
\end{equation}
Unlike other models considered so far, for SIS the mean-field calculation is not
exact in the thermodynamic limit, but captures the overall qualitative behavior:
exactly like in the synchronization case, the inflection in $F(S)$ yields three
possible fixed points, and a saddle node bifurcation involving two of them (see
Fig.~\ref{fig:trans}d).

As before, the same mean-field behavior is obtained with a
graph $\W$ that parametrizes the following infection rate:
\begin{equation}\label{eq:graph_inf}
  \nu_i = \sum_{k}\beta_{k}'\sum_{\cc}\bm 1_{|\cc|=k}\prod_{j\in \cc}W_{ij}x_j(t),
\end{equation}
where $\cc$ corresponds to an arbitrary node subset. If we set
$W_{ij}=\avg{W_{ij}}=z/(N-1)$ and $\beta'_k= \beta_k (N-1)/{N - 1\choose k - 1}$ we recover
the same mean-field of Eq.~\ref{eq:mf_simplicial}. Since in the case $z=O(1)$
the graph is asymptotically locally tree-like, the abrupt transition described
by the mean-field calculation cannot be attributed to the existence of cliques,
let alone hyperedges, or simplices.

\subsection{Stability of ecological systems}\label{sec:stability}

In the context of ecological systems, in Ref.~\cite{bairey_high-order_2016} the
authors consider a population dynamics on a model given by
\begin{equation}
  \dot x_i = -x_i^2 + \sum_j\lambda_{ij} x_ix_j
  + \sum_{jk}\lambda_{ijk} x_ix_jx_k  + \cdots,
\end{equation}
where the entries of $\bl$ for every order are sampled independently from
Gaussian distributions with zero mean and uniform variances. When the variance
of the multivariate terms are changed, the stability of species coexistence also
changes. Since this setup falls squarely within the mean-field ansatz, dynamics
in the thermodynamic limit will remain invariant if the hyperedges are replaced
by graphs, as we discussed previously, so the change in stability is due to the
multiway products in the interactions, rather than particular clique
structures. Similar formulations that occur in other
works~\cite{levine_beyond_2017, gibbs_coexistence_2022}, while not necessarily
employing the same assumptions regarding hypergraph structure, are susceptible
to the same observations.

Another illustrative example is the work of
Ref.~\cite{grilli_higher-order_2017}, where the stability of the following
population dynamics is analyzed:
\begin{equation}
  \dot x_i = -x_i^2 + \sum_{jk}\left[2W_{ij}W_{ik} + W_{ij}W_{jk} + W_{ik}W_{kj}\right]x_ix_jx_k.
\end{equation}
This case is conceptually informative because the model is parameterized
directly by a matrix $\W$ (i.e.\ by a dyadic network representation rather
than a hypergraph). Nevertheless, it is frequently discussed in the HON
literature as a canonical illustration of ``higher-order''
effects~\cite{battiston_networks_2020,battiston_physics_2021,battiston_higher-order_2025},
seemingly in tension with their central claim in equating this term with
hypergraph parameterizations, and highlighting a recurring ambiguity in how
``higher-order'' is used in that literature to mean either multivariate
interactions or the use of hypergraphs.

In fact, it is worth remarking that the authors of
Refs.~\cite{bairey_high-order_2016,grilli_higher-order_2017} neither promote
nor necessarily endorse the ``pairwise'' (graph-based) versus ``higher-order''
(hypergraph-based) dichotomy which is repeatedly advocated in the HON
literature. Rather, they compare specific biological models involving bivariate
and trivariate functions, which are directly relevant in the context of
population dynamics. This distinction is particularly important there because,
unlike the broader field of network science, most traditional ecological models
consider only bivariate interactions, typically of the form in
Eq.~\ref{eq:bivariate}. Numerous studies, including the above, have nevertheless
demonstrated that multivariate interactions can play an important role in
ecological
dynamics~\cite{holling1959,murdoch1969,mayfield_higher-order_2017,lafferty2015,levine_beyond_2017,mickalide_higher-order_2019,sundarraman_higher-order_2020,gibbs_coexistence_2022}.

Although some authors have criticized the recent focus on higher-order
interactions~\cite{aladwani_is_2019}, few ecologists would deny the fundamental
role of basic multivariate interactions that arise from basic competition
effects. A classical example is the case of a predator feeding on multiple prey
species \cite{murdoch1969}. In a food web (a directed graph) this situation is
typically captured by directed links from the prey populations to the predator
population. Although the links thus connect only two populations, predators have
a limited capacity for feeding and thus the presence of one prey species reduces
the predation pressure on the other (apparent competition). The predation of
species 1 on species 2 in the presence of species 3 is then typically modeled by
multivariate functions of the form
\begin{equation}
  P_{2 \to 1} = \frac{ax_1x_2}{b+cx_1+dx_3},
\end{equation}
or generalizations thereof \cite{murdoch1969,gross2009}.

We emphasize that our critique concerns not the relevance of multivariate
interactions \emph{per se}, which we believe to have been clearly demonstrated,
but the common tendency to treat multivariate dependence as synonymous with a
need for hypergraph parameterizations.

\subsection{Special cases of multilayer dynamics}\label{sec:multilayer_dynamics}

In this section we show that a broad class of hypergraph-based dynamical
models---namely those written as order-resolved contributions in node space,
such as Eq.~\ref{eq:hypercrap}, admits equivalent formulations not relying on
hypergraphs. When the dynamics is linear, as in diffusion, classical weighted
dyadic networks are often enough. In nonlinear settings, one can use a
multilayer-network framework, via a “lift” to layers indexed by interaction
order, followed by a projection back to node space. Consequently, hypergraph
structure is not uniquely “maximally general”~\cite{gambuzza_stability_2021}, as
often suggested: the same node-level evolution can be realized by a richer
family of multilayer dynamics. Since this modeling pattern underlies many
higher-order synchronization and stability analyses (e.g.\
Ref.~\cite{mulas2020coupled,gambuzza_stability_2021,gallo2022synchronization}),
the present perspective is directly relevant in that setting.

\subsubsection{Diffusion}

We consider first the case of diffusion, where the state variable represents a
conserved scalar transported across interaction events. Let $x_i\left(t\right)$
denote the amount (or concentration) of a conserved quantity stored at node $i$
at time $t$. Let us suppose that interaction events are specified by a
hypergraph incidence matrix $e_{\cc i}\in\{0,1\}$, where $e_{\cc i}=1$ if node $i$
participates in hyperedge $\cc$, and let $|\cc|=\sum_k e_{\cc k}$ be the hyperedge
size. We consider a linear diffusion rule in which each incident hyperedge $\cc$
provides an exchange channel of unit strength: the content at node $i$ is
drained through each incident hyperedge at rate $1$, and the outflow through $\cc$
is distributed uniformly among the other endpoints of $\cc$. The resulting
mass-balance equation is
\begin{equation}
\label{eq:hyper_diff_EX}
\dot x_i = -x_i\sum_{\cc} e_{{\cc}i} + \sum_{{\cc}}\sum_{j\neq i}\frac{e_{{\cc}i}e_{{\cc}j}}{|{\cc}|-1}x_j,
\end{equation}
which can be written as $\dot{\mathbf{x}}=-\mathbf{L}^{(\mathrm{hyp})}\mathbf{x}$ for a hypergraph-induced generator $\mathbf{L}^{(\mathrm{hyp})}$. However, one can also define a weighted network on the same set of nodes by introducing the “dyadic” weights
\begin{equation}
\label{eq:W_weight_EX}
W^{(\mathrm{dya})}_{ij} = \sum_{\cc} \frac{e_{\cc i}e_{\cc j}}{|\cc|-1} \quad\text{for } i\neq j,
\qquad W^{(\mathrm{dya})}_{ii}=0,
\end{equation}
and the corresponding strength (i.e.\ weighted degree) $k_i^{(\mathrm{dya})} =\sum_j W^{(\mathrm{dya})}_{ij}$. Accordingly, the standard diffusion on this weighted graph reads
\begin{equation}\label{eq:graph_diff_weight_EX}
\dot x_i = -k_i^{(\mathrm{dya)}}x_i + \sum_j W^{(\mathrm{dya)}}_{ij}x_j,
\end{equation}
suggesting that $k_i^{(\mathrm{dya})}=\sum_{\cc} e_{\cc i}$. This simple result shows that hypergraph diffusion is exactly equivalent to diffusion on the weighted graph of~\eqref{eq:W_weight_EX}, without any need to interpret the process as requiring irreducible clique constraints.

Similarly, the same weighted construction can be rewritten as an order-resolved edge-colored graph (i.e.\ a non-interconnected multiplex~\cite{de_domenico_mathematical_2013,kivela2014multilayer,de2023more}) by decomposing the weights by hyperedge size, so that each layer isolates the contribution of interactions of a fixed cardinality. For $m\ge 2$, we define the layer-$m$ weight matrix
\begin{equation}\label{eq:W_layer_m_EX}
W^{(m)}_{ij} = \sum_{\cc:|\cc|=m}\frac{e_{\cc i}e_{\cc j}}{m-1} \quad\text{for } i\neq j, \qquad W^{(m)}_{ii}=0,
\end{equation}
and let $\mathbf{L}^{(m)}$ be the associated weighted Laplacian: accordingly, $\mathbf{L}^{(\mathrm{hyp})}=\sum_{m\ge 2}\mathbf{L}^{(m)}$, so~\eqref{eq:hyper_diff_EX} is recovered either as a single weighted-graph process or as the aggregate of a multilayer process whose layers are indexed by interaction order and whose layer-sum generator equals $\mathbf{L}^{(\mathrm{hyp})}$.

The diffusion case above is deliberately elementary: the equivalence is exact
and follows from matching generators. This happens because the example collapses
to a weighted graph, and we are effectively using a node-space projection of
hyperedge events into pairwise transitions, not a higher-order Laplacian in the
Hodge sense~\cite{lim_hodge_2020}.

Intriguingly, we can apply the same projection-based idea in a setting that is
central in the HON literature, namely the order-resolved variational dynamics
used in higher-order synchronization stability analyses, as we show in the
following.

\subsubsection{Nonlinear dynamics}

A useful way to make the preceding point concrete is to start from a practical
example in which the relevant quantities are already expressed in node space and
the conclusion can be checked by direct substitution. As a particular case of
Eq.~\ref{eq:hypercrap}, we use the order-resolved variational equation around
the synchronous manifold derived in Ref.~\cite{gambuzza_stability_2021}. Under
the standard noninvasiveness condition (the coupling vanishes on synchrony) and
the usual factorization on the synchronous manifold (i.e.\ each interaction
order contributes a node-space operator times an $m\times m$ Jacobian factor),
the linearized dynamics for the stacked perturbation
$\delta\mathbf{x}\in\mathbb{R}^{Nm}$ can be written as
\begin{equation}\label{eq:gambuzza_var_EX}
\delta\dot{\mathbf{x}} = \left[\mathbf{I}_N\otimes \mathbf{J}F - \sum_{d=1}^D \sigma_d \mathbf{\tilde L}^{(d)}\otimes \mathbf{J}G^{(d)}
\right]\delta\mathbf{x} \equiv  \mathbf{A}\delta\mathbf{x},
\end{equation}
where each node has an $m$-dimensional state, so
$\delta\mathbf{x}=\left(\delta\mathbf{x}_1,\ldots,\delta\mathbf{x}_N\right)\in\mathbb{R}^{Nm}$
with $\delta\mathbf{x}_i\in\mathbb{R}^{m}$. Moreover,
$\mathbf{J}F\in\mathbb{R}^{m\times m}$ and
$\mathbf{J}G^{(d)}\in\mathbb{R}^{m\times m}$ are Jacobian factors evaluated on
synchrony, $\mathbf{\tilde L}^{(d)}\in\mathbb{R}^{N\times N}$ is the node-space
operator associated with interactions of order $(d+1)$, and $D$ is the maximum
interaction order included in the decomposition. If $\mathbf{\tilde L}^{(d)}$ is
built from a fully populated symmetric $(d+1)$-body encoding, any $1/d!$
convention can be absorbed either into $\mathbf{\tilde L}^{(d)}$ or into
$\sigma_d$, without affecting the argument below.

Notably, \eqref{eq:gambuzza_var_EX} admits an interpretation in a layer-resolved
manner. To make it explicit, we introduce $D$ copies of the node-space
perturbation, one per interaction order (equivalently, one per layer), and
collect them into the lifted variable
$\delta\mathbf{X}=\left(\delta\mathbf{X}_1,\ldots,\delta\mathbf{X}_D\right)\in\mathbb{R}^{DNm}$,
with $\delta\mathbf{X}_\alpha\in\mathbb{R}^{Nm}$. The projection back to node
space, $\delta\mathbf{x}=\mathbf{P}\delta\mathbf{X}$, is the layer average
\begin{equation}\label{eq:proj_sync_EX}
\mathbf{P}=\frac{1}{D}\left(\mathbf{u}^\top\otimes \mathbf{I}_{Nm}\right),
\end{equation}
where $\mathbf{u}\in\mathbb{R}^{D}$ is the all-ones vector in layer space, that is $\mathbf{u}^\alpha=1$ for $\alpha=1,\ldots,D$ (equivalently, $\mathbf{u}=(1,\ldots,1)^\top$). With this convention one has
\begin{equation}
\left(\mathbf{u}^\top\otimes \mathbf{I}_{Nm}\right)\delta\mathbf{X}=\sum_{\alpha=1}^{D}\delta\mathbf{X}_\alpha
\implies \delta\mathbf{x}=\frac{1}{D}\sum_{\alpha=1}^{D}\delta\mathbf{X}_\alpha.
\end{equation}

Next, let us assign to each layer $\alpha$ an order-resolved operator that
contains a full copy of the local term and only the $\alpha$-th coupling channel
as
\begin{equation}\label{eq:Calpha_sync_EX}
\mathbf{C}^{(\alpha)}=\mathbf{I}_N\otimes \mathbf{J}F - D\sigma_\alpha \mathbf{\tilde L}^{(\alpha)}\otimes \mathbf{J}G^{(\alpha)} \in\mathbb{R}^{Nm\times Nm},
\end{equation}
where the factor $D$ is chosen so that averaging the $D$ layers reproduces the single-copy coefficients of~\eqref{eq:gambuzza_var_EX}. The multiplex dynamics reads
\begin{equation}\label{eq:lift_sync_EX}
\delta\dot{\mathbf{X}}_\alpha = \mathbf{C}^{(\alpha)}\left(\mathbf{P}\delta\mathbf{X}\right) + \sum_{\beta=1}^{D}\mathbf{K}_{\alpha\beta}\delta\mathbf{X}_\beta,
\qquad \alpha=1,\ldots,D,
\end{equation}
where $\mathbf{K}_{\alpha\beta}\in\mathbb{R}^{Nm\times Nm}$ are interlayer
linear maps constrained by the zero-average (gauge) condition
\begin{equation}\label{eq:gauge_sync_EX}
\sum_{\alpha=1}^{D}\mathbf{K}_{\alpha\beta}=\mathbf{0}
\qquad\text{for each fixed }\beta.
\end{equation}

One way to interpret the above construction is as a replicated system with
internal gauge degrees of freedom. The projected perturbation $\delta\mathbf{x}$
is the observable node-level quantity governed by \eqref{eq:gambuzza_var_EX},
while the lifted state
$\delta\mathbf{X}=\left(\delta\mathbf{X}_1,\ldots,\delta\mathbf{X}_D\right)$
replicates this observable into $D$ internal channels (layers) so that each
channel can carry the contribution of a single interaction order. These channels
are not independent copies of the system: they form a decomposition of the same
node-level perturbation, constrained by
$\delta\mathbf{x}=\mathbf{P}\delta\mathbf{X}=D^{-1}\sum_{\alpha=1}^{D}\delta\mathbf{X}_\alpha$.
For this reason it is natural that the order-resolved operator
$\mathbf{C}^{(\alpha)}$ acts on the common input $\delta\mathbf{x}$ rather than
on $\delta\mathbf{X}_\alpha$ itself. The interlayer term $\mathbf{K}$ then
governs how the internal channels exchange degrees of freedom among
themselves without changing the mean, which is precisely what the gauge
condition~\eqref{eq:gauge_sync_EX} enforces.

Accordingly, projecting~\eqref{eq:lift_sync_EX} with~\eqref{eq:proj_sync_EX} gives
\begin{equation}\label{eq:proj_calc_sync_EX}
\delta\dot{\mathbf{x}} = \mathbf{P}\delta\dot{\mathbf{X}} = \frac{1}{D}\sum_{\alpha=1}^{D}\mathbf{C}^{(\alpha)}\delta\mathbf{x} + \frac{1}{D}\sum_{\beta=1}^{D}\left(\sum_{\alpha=1}^{D}\mathbf{K}_{\alpha\beta}\right)\delta\mathbf{X}_\beta,
\end{equation}
where the second term vanishes by~\eqref{eq:gauge_sync_EX}. Using~\eqref{eq:Calpha_sync_EX}, we finally obtain
\begin{equation}\label{eq:avgC_sync_EX}
\frac{1}{D}\sum_{\alpha=1}^{D}\mathbf{C}^{(\alpha)} =
\mathbf{I}_N\otimes \mathbf{J}F - \sum_{\alpha=1}^{D}\sigma_\alpha \mathbf{\tilde L}^{(\alpha)}\otimes \mathbf{J}G^{(\alpha)} = \mathbf{A},
\end{equation}
so~\eqref{eq:proj_calc_sync_EX} reduces to
$\delta\dot{\mathbf{x}}=\mathbf{A}\delta\mathbf{x}$, i.e.\ exactly like
\eqref{eq:gambuzza_var_EX}. Since~\eqref{eq:gauge_sync_EX} admits infinitely
many choices of $\left\{\mathbf{K}_{\alpha\beta}\right\}$, the same projected
node-space evolution is compatible with infinitely many realizations of
multiplex dynamics. In this sense, \eqref{eq:gambuzza_var_EX} by itself does not
identify a unique microscopic interpretation in terms of ``higher-order'' versus
``layered'' structure.

Having established the mechanism on this explicit, widely used case study, the
same construction extends to a broad class of order-decomposable hypergraph
dynamics of the form~\eqref{eq:hypercrap}. We write~\eqref{eq:hypercrap}
compactly as
\begin{equation}\label{eq:hyper_compact_EX}
  \dot{\mathbf{x}}=\mathbf{F}\left(\mathbf{x}\right)+\sum_{d=1}^{D}\mathbf{H}^{(d)}\left(\mathbf{x}\right),
\end{equation}
with
\begin{equation}
\qquad \left[\mathbf{H}^{(d)}\left(\mathbf{x}\right)\right]_i:=\sum_{j_1,\ldots,j_d=1}^N \lambda_{i j_1\cdots j_d}\mathbf{f}\left(\mathbf{x}_i,\mathbf{x}_{j_1},\ldots,\mathbf{x}_{j_d}\right).
\end{equation}
We define the same $D$-layer state $\mathbf{X}\in\mathbb{R}^{DNm}$ and
projection $\mathbf{x}=\mathbf{P}\mathbf{X}$ with $\mathbf{P}$ as
in~\eqref{eq:proj_sync_EX}. We now consider the lifted nonlinear multiplex dynamics
\begin{multline}\label{eq:lift_hyper_EX}
\dot{\mathbf{X}}_\alpha = \mathbf{F}\left(\mathbf{P}\mathbf{X}\right)
+ D\mathbf{H}^{(\alpha)}\left(\mathbf{P}\mathbf{X}\right)
+ \sum_{\beta=1}^{D}\mathbf{K}_{\alpha\beta}\left(\mathbf{X}\right),\\
\alpha=1,\ldots,D,
\end{multline}
where the interlayer terms satisfy the pointwise zero-average constraint
\begin{equation}\label{eq:gauge_hyper_EX}
\sum_{\alpha=1}^{D}\mathbf{K}_{\alpha\beta}\left(\mathbf{X}\right)=\mathbf{0},
\qquad\text{for all }\mathbf{X}\text{ and each fixed }\beta.
\end{equation}
Finally, projecting~\eqref{eq:lift_hyper_EX} with $\mathbf{P}$ leads to the equation
\begin{equation}\label{eq:proj_hyper_EX}
\dot{\mathbf{x}} = \mathbf{P}\dot{\mathbf{X}} = \mathbf{F}\left(\mathbf{x}\right) + \sum_{\alpha=1}^{D}\mathbf{H}^{(\alpha)}\left(\mathbf{x}\right) + \frac{1}{D}\sum_{\beta=1}^{D}\left(\sum_{\alpha=1}^{D}\mathbf{K}_{\alpha\beta}\left(\mathbf{X}\right)\right),
\end{equation}
and the last term vanishes by~\eqref{eq:gauge_hyper_EX}, giving
$\dot{\mathbf{x}}=\mathbf{F}\left(\mathbf{x}\right)+\sum_{d=1}^{D}\mathbf{H}^{(d)}\left(\mathbf{x}\right)$,
which is exactly~\eqref{eq:hypercrap}. As in the synchronization example, the
admissible family of $\left\{\mathbf{K}_{\alpha\beta}\right\}$ is infinite, so
the same node-level hypergraph dynamics corresponds to infinitely many multiplex
realizations under the layer-average projection.

\section{Lack of empirical grounding}\label{sec:empirical}

Both networks and hypergraphs are not objects of the real world but conceptual
abstractions used to model complex systems. A single real-world system can
sometimes be represented in many different ways---as a network, as a hypergraph,
or through entirely different mathematical frameworks. The choice of
representation is therefore guided by both preference and utility. In practice,
it requires balancing mathematical versatility, interpretability, parsimony, and
empirical support~\cite{peel_statistical_2022}.

Historically, network-based models have proven extremely powerful. Representing
a system as a network is a strong simplification, yet it enables the application
of a rich set of analytical tools. These include spectral
methods~\cite{chung_spectral_1997}, the analysis of degree distributions and
associated generating-function techniques~\cite{newman_random_2001}, and the
formulation of graph
ensembles~\cite{erdos_random_1959,erdos_graphs_1960,chung_average_2002,holland_stochastic_1983},
among many others~\cite{newman_networks:_2010}.

As discussed above, hypergraph parameterizations impose more restrictive
mathematical structures on large-scale interacting systems, whereas graph-based
formulations tend to offer greater generality. In addition to this increased
structural specificity, hypergraphs currently come with a more limited set of
mature analytical tools. Although recent advances have extended ensemble
approaches to
hypergraphs~\cite{young_construction_2017,chodrow_configuration_2020,chodrow_generative_2021,preti_higher-order_2024,saracco_entropy-based_2025},
these methods are not yet as developed or as broadly applicable as classical
graph models such as the \ER random graph~\cite{erdos_random_1959}, the
configuration model~\cite{chung_average_2002}, and their
generalizations~\cite{holland_stochastic_1983, karrer_stochastic_2011}.
Likewise, many elegant and general results in graph theory, such as Euler’s
solution to the K\"{o}nigsberg bridge problem~\cite{euler_solutio_1741} or
generating-function approaches to giant-component
analysis~\cite{newman_random_2001}, do not yet have comparable counterparts in
hypergraph theory, some specialized frameworks
notwithstanding~\cite{zlatic_hypergraph_2009,bianconi_theory_2024}. In
particular, tensor-based formulations of hypergraphs have so far yielded limited
progress toward a broadly useful spectral theory~\cite{shetty_spectral_2025}.
Future developments may well change this situation, and hypergraphs could become
an increasingly powerful modeling framework. At present, however, in the absence
of clear and consistent mathematical advantages, networks (including multilayer
networks) often provide a more practical framework for many applications.

From a strictly empirical perspective, however, mathematical tractability is not
the primary consideration. Because of their greater structural and functional
specificity, hypergraph models may offer more parsimonious representations of
systems characterized by overlapping groups of nodes engaged in symmetrical,
reciprocal interactions, as we discuss in Sec.~\ref{sec:mdl}. Consequently, the
choice between graph and hypergraph descriptions is inherently empirical and
context dependent, and should ultimately be guided by evidence about which
representation captures the relevant structure of the system most
effectively---keeping in mind that due to the expressiveness hierarchy we
discussed previously, many graph-based models simply do not admit
hypergraph-based formulations.

Despite this, claims regarding the ubiquity and general suitability of
hypergraph models are often supported by arguments that, we contend, do not
constitute meaningful empirical evidence. These arguments typically take one of
the following forms:
\begin{enumerate}
\item The existence of toy models of network behavior formulated using hypergraphs.
\item The reinterpretation of bipartite network data as hypergraph data.
\item The imputation of hypergraphs from graph data.
\item The heuristic inference of hypergraphs from time-series data.
\end{enumerate}

We emphasize we are not addressing the empirical evidence for multivariate
interactions, which is
substantial~\cite{centola_spread_2010,monsted_evidence_2017,mickalide_higher-order_2019,sundarraman_higher-order_2020},
but instead for their specific connection with hypergraph structures.

In what follows, we examine each of these classes of purported evidence in turn.
We conclude with a general explanation for the generalized absence of evidence
for hypergraph formulations.

\subsection{Toy models are not evidence}

Claims regarding the purported ubiquity of hypergraphs in nature are frequently
justified by the existence of simplified mathematical models proposed for
specific
phenomena~\cite{battiston_networks_2020,bianconi_higher-order_2021,torres_why_2021,battiston_physics_2021,bick_what_2023,battiston_higher-order_2025,abiad_hypergraphs_2026}.
Examples include models of synchronization~\cite{skardal_higher_2020},
population dynamics~\cite{burgio_evolution_2020,
  alvarez-rodriguez_evolutionary_2021, wang_evolutionary_2024,
  civilini_explosive_2024}, epidemic spreading~\cite{matamalas_abrupt_2020}, and
equilibrium spin systems~\cite{son_phase_2024, robiglio_higher-order_2025},
among others.

In these cases, stylized models are constructed and analyzed to reproduce
particular behaviors, yet no empirical justification is provided for their
suitability. Typically, the choice of a hypergraph formulation is motivated by
a subjective conceptual appeal rather than by direct evidence that the underlying
system exhibits this particular kind of multivariate interactions. Even setting
aside the fact---discussed above---that each such model represents a special
case of a more general graph-based formulation, and that the same qualitative
behaviors can often be reproduced using suitably structured graph-based
interactions (e.g.\ locally tree-like or multilayer graphs), the mere existence
of these models cannot substitute for empirical evidence.

More fundamentally, qualitatively reproducing a macroscopic phenomenon within a
hypergraph model does not establish the necessity, or even the relevance, of
hypergraph structure for the system in question. Without independent direct
evidence demonstrating the particular role of hypergraph-based interactions,
such models remain underdetermined, since multiple, non-equivalent microscopic
descriptions can give rise to the same observed macroscopic behavior. In this
context, invoking hypergraphs risks conflating the role of multivariate
functions with network structure, particularly when no systematic model
selection, falsification, or comparison against alternative graph-based models
is performed.

Consequently, one cannot plausibly claim that hypergraphs are ``ubiquitous'' or
``essential'' to model nature just because many toy models using them appear in
the literature. At best, these models demonstrate that hypergraphs are a
possible modeling choice; they do not demonstrate that they are empirically
warranted, uniquely informative, or required to capture the phenomena under
study.

\subsection{Bipartite networks reimagined}

This class of claims consists of reinterpreting data that have traditionally
been analyzed as bipartite graphs as hypergraphs. Such claims generally fall
into two categories: interpretational alternatives and modeling alternatives.

Interpretational claims are well illustrated by co-authorship
networks~\cite{patania_shape_2017, battiston_higher-order_2025}, where a
bipartite graph connecting authors and papers is reimagined as a hypergraph in
which each paper becomes a hyperedge representing an abstract ``interaction''
among authors.\footnote{Since these ``interactions'' are left unspecified, it
  would appear equally valid to model authors as ``interactions'' between
  papers---yet this alternative is never pursued.} A similar reinterpretation is
applied to spatiotemporal proximity data, where nodes appearing within the same
time window are grouped into a
hyperedge~\cite{iacopini_temporal_2024,battiston_higher-order_2025,
  mancastroppa_adaptive_2026}. Because bipartite graphs and hypergraphs are
mathematically equivalent objects,\footnote{We note in passing that all recent
  software packages for hypergraphs that service the HON literature rely
  internally on bipartite adjacency lists~\cite{landry_xgi_2023,
    lotito_hypergraphx_2023, praggastis_hypernetx_2024}. Since these data
  structures are identical to those employed in general graph-based frameworks,
  such software offers no inherent representational advantage, beyond providing
  a tailored set of algorithms that could equally well be implemented within
  more general frameworks.} this practice often amounts to a
relabeling exercise. For example, degree distributions are replaced by hyperedge
order distributions, yielding mathematically identical descriptions under
different terminology~\cite{battiston_higher-order_2025}.

Although hypergraph language can occasionally foreground concepts that are less
commonly discussed using bipartite representations~\cite{patania_shape_2017},
this distinction is ultimately terminological rather than empirical. Setting
aside the limited epistemic value of representing a paper as a monolithic,
indivisible ``interaction'' among
authors~\cite{battiston_higher-order_2025}---rather than as the outcome of a
complex, internally structured process\footnote{Large paper collaborations and
  consortia---often involving tens, hundreds, or even thousands of authors---are
  virtually always divided into
  subteams~\cite{atlas_collaboration_combined_2015}. This structure is
  frequently evident not only in the text of the paper itself, but also in
  metadata such as affiliations and departmental subdivisions. These subteams
  likely possess their own internal formal and informal substructures, and are
  embedded within an overall pattern of sparse interactions, with a large
  contingent of co-authors who have not interacted meaningfully, or may not have
  met each other at all. Even in small collaborations, tasks and effort are
  rarely evenly distributed, and individual author contributions are typically
  heterogeneous. Models that explicitly account for such internal
  structures~\cite{do_patterns_2010} therefore seem far more plausible than
  indivisible, structureless group interactions, as advocated, for example, in
  Ref.~\cite{battiston_higher-order_2025}}---and making similar considerations
for spatiotemporal proximity data, the mere replacement of terminology provides
no substantive evidence for the suitability of hypergraph formulations.

The second subclass consists of generative models and algorithms designed to
extract structure from bipartite data using hypergraph parameterizations. A
prominent example is community detection, where hypergraph formulations are
employed to cluster nodes either in graph data~\cite{benson_higher-order_2016}
or within a single mode of bipartite networks~\cite{chodrow_generative_2021,
  contisciani_inference_2022, sales-pardo_hyperedge_2023}. This line of inquiry
is, in principle, entirely legitimate: as noted earlier, hypergraph models may
yield more parsimonious descriptions in specific cases.

In practice, however, many such approaches rely on heuristic descriptions that
do not reveal statistical
evidence~\cite{benson_higher-order_2016,carletti_random_2021,kaminski_modularity_2024,kovacs_community_2025},
and therefore do not permit meaningful comparisons with graph-based
alternatives~\cite{peel_statistical_2022, peixoto_descriptive_2023,
  cimini2019statistical}. A smaller subset of works represents early steps
toward inferential approaches based on explicit generative
models~\cite{chodrow_generative_2021, contisciani_inference_2022,
  sales-pardo_hyperedge_2023}. These studies, however, provide little or no
systematic comparison with alternative models for
graphs~\cite{peixoto_disentangling_2022} or bipartite
networks~\cite{larremore_efficiently_2014, gerlach_network_2018,
  yen_community_2020}, nor do they employ robust model selection techniques such
as Bayesian nonparametric inference or the minimum description length
principle~\cite{peixoto_bayesian_2019, peixoto_descriptive_2023}, a topic we
revisit in Sec.~\ref{sec:mdl}.

As it stands, there is no substantial empirical evidence that these
hypergraph-based models provide superior descriptions of bipartite network data
compared to models that do not invoke hypergraph formulations.

\subsection{Imputation from graph data}

The third class of empirical analyses involves extracting hypergraph structure
from graph data, treating the observed graph as a projection of an underlying
latent hypergraph~\cite{patania_shape_2017, battiston_higher-order_2025}. In
principle, this problem can be formulated as a statistical inference task and
addressed using explicit generative models and principled inferential
methods~\cite{chodrow_generative_2021, contisciani_inference_2022}. In practice,
however, it is more commonly approached heuristically: cliques in the graph are
directly interpreted as hyperedges~\cite{benson_higher-order_2016,
  patania_shape_2017, iacopini_simplicial_2019, skardal_higher_2020}, or groups
of pairwise proximity events that occur inside a specific time interval are
declared as
hyperedges~\cite{iacopini_simplicial_2019,cencetti_temporal_2021,myers_topological_2023,lee_temporal_2023}.

Beyond the obvious issue of statistical significance---namely, that dense
subgraphs and cliques can arise purely from stochastic fluctuations or
lower-order correlations---this heuristic identification suffers from deeper
conceptual limitations. First, the inverse problem is fundamentally
non-identifiable: many distinct hypergraph configurations can generate the same
projected graph, even under restrictive
assumptions~\cite{young_hypergraph_2021}. Without a well-specified generative
model and an explicit model selection criterion, there is no principled basis
for preferring one inferred hypergraph over another, rendering the extracted
hypergraph structure largely arbitrary.

Second, and perhaps more fundamentally, this approach undermines any claim that
hypergraph representations capture supposedly irreducible ``higher-order
interactions.'' If hyperedges are defined precisely as maximal cliques in the
observed graph, then any relevant higher-order structure is, by construction,
reducible to a superposition of edges. In this setting, hypergraphs do not
reveal patterns beyond those already encoded in the graph; they merely repackage
graph-theoretic features in hypergraph language.

Based on these considerations we argue that clique-based extraction of
hypergraph structure from graphs neither establishes the empirical necessity of
hypergraph representations nor provides evidence for organization principles
that go beyond what can already be accounted for using conventional graph models.

\subsection{Heuristic inference from time series}

Finally, the last class of claims concerns the inference of hypergraph models
from indirect data, such as time
series~\cite{kralemann_reconstructing_2011,pikovsky_reconstruction_2018,rosas_quantifying_2019,wang_full_2022,santoro_higher-order_2023,malizia_reconstructing_2024,santoro_higher-order_2024,delabays_hypergraph_2025}.
Some works in this category develop algorithms tailored to specific hypergraph
models~\cite{kralemann_reconstructing_2011,pikovsky_reconstruction_2018,wang_full_2022,malizia_reconstructing_2024,su_distinguishing_2025},
assuming that the data are generated by the model and measured perfectly.
Because real empirical data are typically noisy, incomplete, and produced by
partially unknown mechanisms, such idealized assumptions limit the applicability
of these methods to practical inference settings and reduce their robustness to
model misspecification. Moreover, these approaches generally do not address
model selection---that is, whether hypergraph models provide more parsimonious
descriptions of empirical data---and thus implicitly assume the correctness of
the hypergraph representation rather than testing it against alternatives.
Without explicit model comparison, reconstruction alone cannot establish whether
the added complexity of hypergraph parameterizations is empirically warranted.
Furthermore, many of these studies are purely theoretical, with no application to
real
datasets~\cite{kralemann_reconstructing_2011,pikovsky_reconstruction_2018,malizia_reconstructing_2024}.

Other analyses infer hypergraph structure from time-series
correlations~\cite{yu_higher-order_2011,petri_homological_2014,rosas_quantifying_2019,herzog_genuine_2022,expert_higher-order_2022,santoro_higher-order_2023,santoro_higher-order_2024,dolgov_higher-order_2023,li_higher-order_2025,camino-pontes_brain_2025,santoro_beyond_2025}.
Such correlation-based methods are notoriously unreliable at distinguishing
direct from indirect interactions, often producing distortions such as
artificially high triangle counts or spurious community
structure~\cite{peel_statistical_2022}. This limitation is particularly severe
in higher-order settings, where indirect pairwise correlations can masquerade as
multibody interactions and lead to systematic overestimation of multivariate
relationships. Compounding this issue, many such approaches do not differentiate
between hypergraph parameterizations and more general multivariate interactions,
further limiting the conclusions that can be drawn from them. Critically, these
approaches do not incorporate empirical evidence in a statistically principled
framework---one that would require explicit likelihood models, complexity
penalties, and formal comparison between competing hypotheses, as we describe in
Sec.~\ref{sec:mdl}---and thus cannot resolve the model selection problem. Taken
together, these limitations prevent existing reconstruction methods from
establishing whether hypergraph structure is genuinely supported by empirical
data, as opposed to being an artifact of modeling assumptions. To date, no
reconstruction method has been developed---or successfully applied to empirical
data---that properly accounts for model
complexity~\cite{peel_statistical_2022,peixoto_network_2019,peixoto_network_2025}
and demonstrates that hypergraph parameterizations yield more plausible
representations for specific systems.

\subsection{Lack of evidence is not incidental}

The lack of empirical support for hypergraph models has been noted in the
higher-order network (HON) literature~\cite{battiston_physics_2021,
  ferraz_de_arruda_contagion_2024}. We argue that this absence of evidence is
not incidental: it reflects a deeper epistemic issue. In short, there is no
interacting system that is representable as a hypergraph but not as a graph
(although the converse is often true); thus, attempts to identify such systems
are likely to be futile. Differences between these network models arise from the
specific rules governing interactions, not from general structural properties of
the system.

Network data are often incomplete, and measurements vary in
accuracy~\cite{newman_network_2018}. However, the central issue here goes beyond
this: the underlying rules governing interactions are almost always
\emph{latent}. For instance, a protein–protein interaction database typically
indicates only whether the interaction is believed to be positive or negative; a
survey reporting friendship ties between students does not specify how gossip or
infection would propagate; and a neuronal map does not reveal firing dynamics.
This pattern holds for virtually all network data. Because distinctions between
graph- and hypergraph-based models—or any other parameterization—depend partly
on these latent rules, direct evidence favoring one representation over another
may never exist.\footnote{Although one could easily rule out the applicability
  of hypergraph-based models if the interaction structure does not possess any
  cliques.} This is standard in science: models are not verified directly, but
are evaluated comparatively, with the most appropriate representation prevailing
based on parsimony, predictive performance, and explanatory
power~\cite{peel_statistical_2022}.

\section{Expressiveness, parsimony, and model selection}\label{sec:mdl}

As we have demonstrated,
hypergraph-based models are not generalizations of graph-based models, but in
fact special cases of them. This means that there are many graph-based dynamical
models that simply cannot be expressed as hypergraphs, which changes the
expected outcome of model selection in important ways. Firstly, if we consider
the set of all possible dynamical systems, only a vanishingly small minority of
them admit a hypergraph-based description, let alone one that is more
parsimonious. Secondly, the hypergraph description can only more parsimonious 
in the special case where the nodes of each hyperedge interact symmetrically 
and mutually. In the relatively few cases where both
representations are simultaneously admissible, parsimony offers no objective
criterion for distinguishing between them, since there is always a graph-based
representation with at most equal complexity.

The HON literature occasionally recognizes, perhaps most prominently in
Ref.~\cite{battiston_physics_2021}, that hypergraph descriptions should be
adopted only to the extent that the empirical evidence available in the data
supports them. The argument typically put forth is that the purported increased
expressiveness of hypergraph-based models, compared to graph-based ones, comes
at the cost of an inherently increased model complexity, which demands more
statistical evidence according to the principle of maximum parsimony. Ignoring
this principle could lead to overfitting, i.e., to mistaking statistical
fluctuations in the data for meaningful signals, resulting in overly complex
models that deviate from reality.

The principle of maximum parsimony is not only sound, but in fact indispensable
when multiple hypotheses need to be compared, as is virtually always the case in
empirical scenarios. In network science, the need for robust model selection
under this principle has been explicitly
advocated~\cite{peel_statistical_2022}, and used extensively, for example, in
the context of community
detection~\cite{rosvall_information-theoretic_2007,hofman_bayesian_2008,guimera_missing_2009,peixoto_parsimonious_2013,peixoto_bayesian_2019,peixoto_hierarchical_2014,peixoto_model_2015,come_model_2015,peixoto_descriptive_2023,peixoto_implicit_2023},
network
reconstruction~\cite{peixoto_network_2019,young_hypergraph_2021,young_bayesian_2021,peixoto_network_2025},
and other instances of model
selection~\cite{kirkley_compressing_2023,kirkley_inference_2024,hebert-dufresne_network_2024},
under the equivalent formalisms of Bayesian inference and the minimum
description length (MDL)
principle~\cite{rissanen_modeling_1978,grunwald_minimum_2007}. Indeed, in
Sec.~\ref{sec:empirical}, we criticized purported empirical evidence
for hypergraph models for failing to incorporate this essential statistical
criterion.

In the following, we elaborate on the problem of Bayesian model selection---or,
equivalently, on the use of the minimum description length principle---and on
how it pertains to the choice between graph- and hypergraph-based descriptions.

In the context of complex interacting systems, given some observation $\D$ (e.g.
a time-series\footnote{In the following, without loss of generality since no
  measurement can be made with infinite precision, $\D$ is assumed to be
  discrete, so that it can be modeled with a probability mass function, and
  hence can be encoded with a finite amount of information.}), and two competing
modelling choices $\mathcal{M}_1$ and $\mathcal{M}_2$, we judge their relative
plausibility according to the posterior odds ratio\footnote{Note that
  Eq.~\ref{eq:podds} does not necessarily assume that the dynamics is
  stochastic. For deterministic systems, measured with negligible noise, we
  would have simply $P(\D|\mathcal{M}) = 1$ for $\D$ the deterministic dynamics
  that evolves from $\mathcal{M}$, and $0$ otherwise.}
\begin{align}\label{eq:podds}
  \frac{P(\mathcal{M}_1|\D)}{P(\mathcal{M}_2|\D)} &= \frac{P(\D|\mathcal{M}_1)P(\mathcal{M}_1)}{P(\D|\mathcal{M}_2)P(\mathcal{M}_2)}\\
  &= 2^{\Sigma(\D,\mathcal{M}_2) - \Sigma(\D,\mathcal{M}_1)},
\end{align}
which is directly related to the respective description lengths~\cite{rissanen_modeling_1978,grunwald_minimum_2007}, i.e.
\begin{equation}
  \Sigma(\D,\mathcal{M}_1) = -\log_2P(\D|\mathcal{M}_1)-\log_2P(\mathcal{M}_1),
\end{equation}
and likewise for $\mathcal{M}_2$, where the first term on the right-hand side
quantifies the amount of information required to encode the data when the model
is known, while the second accounts for the amount of information required to
encode the model itself. The above means that the most plausible model is the
one that allows for the most compression of the data, and hence that the
complexity of a model needs to be quantified before its plausibility can be
assessed.

In the following, we will consider an arbitrary graph-based model
$\mathcal{M}_G=(\f, \W)$ and an arbitrary hypergraph-based model
$\mathcal{M}_H=(\g, \bm \lambda)$, according to our general framework of
Eq.~\ref{eq:netdyn} and Eq.~\ref{eq:hyperdyn}, which leads to the model terms of
the description length,
\begin{align}
  -\log_2P(\mathcal{M}_G) &= -\log_2P(\f | \W) - \log_2P(\W)\\
  -\log_2P(\mathcal{M}_H) &= -\log_2P(\g | \bm \lambda ) - \log_2P(\bm \lambda),
\end{align}
where $P(\W)$ and $P(\bm \lambda)$ are generative models that encode the graph
and the hypergraph, respectively, and likewise $P(\f | \W)$ and
$P(\g | \bm \lambda)$ are generative models for graph-based and hypergraph-based
interaction functions, $\f$ and $\g$ respectively, that take their respective
inputs as parameters.

To make our analysis more concrete,
following~\cite{battiston_networks_2020,bianconi_higher-order_2021,torres_why_2021,battiston_physics_2021,majhi_dynamics_2022,bick_what_2023,battiston_higher-order_2025,abiad_hypergraphs_2026}
we consider, as an example, $\g$ to be admissible if it corresponds to
compositions of individual multivariate functions $h_{\cc}$ on each incident
hyperedge $\cc$, i.e.
\begin{equation}\label{eq:hyperg}
  g_i(x_i, \{ \bm{x}_{\cc \setminus i} \mid \lambda_{\cc} =1, i\in\cc\})
  = \sum_{\cc}\bm{1}_{i\in \cc}\lambda_{\cc}h_{\cc}(\bm x_{\cc}),
\end{equation}
where one could also replace the sum by any other elementary operation, without
any substantial consequence for the ensuing analysis.\footnote{Although the definition of
  Eq.~\ref{eq:hyperg} covers what is typically considered in the HON
  literature, our conclusions remain unchanged also for
  generalizations, for example when the members of a particular hyperedge
  experience the interactions differently. The crux of our argument, as
  discussed earlier, is that hyperedge parametrizations must enforce some form
  of additional constraint on the interaction functions, constraints which are
  absent in a graph-based setting. The particular form these extra constraints
  take is not of primary importance for our present analysis.}

For what follows, it is essential to keep in mind the non-invertible mapping
between these model classes, whereby every choice of $(\bm \lambda, \g)$ can be
non-injectively mapped to a $(\f, \W)$ description given by
\begin{equation}
  f_i(x_i, \{x_j \mid W_{ij} = 1\}) =  g_i(x_i, \{ \bm{x}_{\cc \setminus i} \mid \lambda_{\cc} =1, i\in\cc\})\label{eq:fproj}
\end{equation}
and
\begin{equation}
  W_{ij} = \begin{cases}1 & \text{ if } \sum_{\cc}\bm{1}_{i \in \cc}\bm{1}_{j \in \cc}\lambda_{\cc} > 0\\ 0 &\text{ otherwise,}\end{cases}\label{eq:wproj}
\end{equation} 
while the converse is not necessarily true, i.e. not every choice of $(\f, \W)$
can be decomposed into some $(\g, \bl)$. It is important to emphasize that the
above non-invertible mapping means that hypergraph-based models represent a
strict special case, both functionally and structurally. The latter property is
immediate: the projection of a hypergraph leads to the existence of cliques,
which need not be present in an arbitrary graph (see
Fig.~\ref{fig:graph-comp}b); hence any graph $\W$ without loops of length three
cannot be represented by any hypergraph $\bl$, since Eq.~\ref{eq:wproj} cannot
be fulfilled, regardless of how we choose the functions $\g$. But as we will now
demonstrate, even for graphs that can accommodate projections of hypergraphs,
graph-based models are still functionally more general.

Let us consider discrete multivariate functions $X^k \to X$, with
$|X|=A \in \mathbb N$, then the number of possible functions with $k$ arguments
is given by $\Omega(A, k)= A^{A^k}$\footnote{The enumeration
  $\Omega(A,k) = A^{A^k}$ also counts functions where some of the arguments are
  ``insensitive'', i.e. they do not affect the output of the function. According
  to the argument of footnote \ref{truism}, these inputs should be omitted from
  the network description. The number of functions with $k$ \emph{sensitive}
  inputs is given instead by
  $\Omega'(A,k) = \sum_{i=0}^k (-1)^i {k\choose i}A^{A^{k-i}}$. Note, however,
  that $\Omega(A,k)/\Omega'(A,k)= 1 - O(k/A^{A^{k-1}})$, meaning that both
  numbers become asymptotically close super-exponentially in $k$, so we use the
  simpler expression instead, without sacrificing anything important.}, sampled
uniformly at random. Then we have a description length contribution given by
\begin{equation}
  -\log_2P(\f | \W) = \sum_i\log_2\Omega(A, d_i + 1) =\sum_iA^{d_i+1}\log_2 A,
\end{equation}
where $d_i=\sum_jW_{ij}$ is the degree of node $i$.
For a hypergraph-based model we have instead
\begin{align}
  -\log_2P(\g | \bl) &= \sum_{\cc}\lambda_{\cc}A^{|\cc|}\log_2 A,\\
  &=\sum_i\sum_{\cc}\bm{1}_{i\in\cc}\frac{\lambda_{\cc}}{|\cc|}A^{|\cc|}\log_2 A.
\end{align}

If we consider the situation, typical for sparse hypergraphs, where the
hyperedges incident on the same node do not share any of their remaining
endpoints, and consider its projected graph $\W$ that fulfills
Eq.~\ref{eq:wproj}, we have that its node degrees are given by
\begin{equation}
  d_i = \sum_{\cc}\bm{1}_{i\in\cc}\lambda_{\cc}(|\cc| - 1),
\end{equation}
from which it follows that
\begin{equation}\label{eq:fgcomp}
  -\log_2P(\f | \W) \ge -\log_2P(\g | \bl),
\end{equation}
with the equality holding only for the trivial case of an empty projected
graph.\footnote{Eq.~\ref{eq:fgcomp} follows from
  $A^{1 + \sum_{j=1}^{m} (k_j - 1)} > \sum_{j=1}^{m} A ^ {k_j}/k_j$ being true
  for $A \ge 2$ and $k_j > 1$.} This fact has the following implications:
\begin{enumerate}
  \item Since the number of graph-based functions $\f$ is strictly larger than
        the number of hypergraph-based functions $\g$, for $\W$ and $\bl$
        compatible according to Eq.~\ref{eq:wproj}, no injective mapping from
        $\f$ to $\g$ can exist.
  \item Due to the above, some choices of $\f$ are simply not representable by
        any $\g$, even for hypergraphs that project to the same graph. For
        these choices, the graph-based model is infinitely more compressible
        than any hypergraph-based alternative.
  \item Conversely, any admissible $\g$ will be more compressible than
        its equivalent $\f$, for the same projected graph $\W$, when a
        uniform encoding for $\bm f$ is used (which means that this encoding is
        not optimal; see the discussion below relating to Eq.~\ref{eq:fcompress}).
\end{enumerate}
Point 3 above demonstrates how generative models tailored for hypergraph-based
descriptions can provide more parsimonious representations in the particular
cases where they are admissible (both structurally and functionally), when
compared to graph-based models that are more general. On the other hand, there
is no general equivalence: the higher expressive power of graph-based
formulations means that they are more generically applicable, and hence
compressive, in situations where hypergraphs are simply not admissible.

A few further considerations are worth making. Firstly, in the setting above, it
is possible for hypergraph-based models to overfit, even when they are
admissible. This happens when hyperedges overlap sufficiently. For example, for
a complete hypergraph where every possible hyperedge from order $2$ to $N$
exists, we have simply
\begin{align}
  -\log_2P(\g | \bl) &= \sum_{k=2}^{N}{N\choose k}A^{k}\log_2 A,\\
  -\log_2P(\f | \W) &= NA^{N}\log_2 A,
\end{align}
and hence $-\log_2P(\g | \bl) > -\log_2P(\f | \W)$ for sufficiently large
$N$,\footnote{This follows from the binomial theorem,
  $\sum_{k=2}^N {N\choose k} A^k = (1+A)^N - NA > NA^N$, for sufficiently large
  $N$.} and thus the graph-based description is compressive, despite the
hypergraph-based one being admissible. Note that this does not mean that the
hypergraph-based model is more general---we can still map it to a graph-based
model using Eqs.~\ref{eq:fproj} and~\ref{eq:wproj} as always. It only means that
in this case the hypergraph-based model is over-parametrized. Similar results
can be obtained for less redundant cases, where not every possible hyperedge
exists.

Secondly, in most empirical settings, one is not interested in uniform encodings
for interaction functions, since these are usually confined to very specific
classes---such as threshold and majority functions, functions with input
symmetry~\cite{gleeson_binary-state_2013}, canalyzing functions, etc.---which
require tailored generative models that provide superior encodings for them. The
constrained set of such functions is typically much smaller, and their
description length contribution scales differently with the number of arguments
than for the uniform encodings considered above. But in general this does not
change the expressiveness hierarchy between graph- and hypergraph-based models,
as long as the sub-families being considered are sufficiently general, nor the
consequences it has for the outcomes of model selection.

Thirdly, every hypergraph-based generative model $P(\g,\bl)$ can be converted
into a graph-based generative model via marginalization, i.e.
\begin{equation}\label{eq:gmarginal}
P'(\f,\W) = \sum_{\g,\bl}\delta_{\f,\f(\g)}\delta_{\W,\W(\bl)}P(\g,\bl),
\end{equation}
where $\f(\g)$ and $\W(\bl)$ represent the solutions of Eq.~\ref{eq:fproj}
and~\ref{eq:wproj}, respectively. Therefore, for this particular joint
distribution (which is in general not compatible with the uniform $P(\f|\W)$
considered earlier), taking the logarithm of both sides of the above equation
and employing Jensen's inequality leads to
\begin{equation}\label{eq:fcompress}
-\log_2P'(\f,\W) \leq -\log_2P(\g,\bl),
\end{equation}
for any pair of compatible representations, with equality achieved only when
there is a single hypergraph-based model $(\g,\bl)$ compatible with a given
graph-based model $(\f,\W)$. In this case, the graph-based description is simply
a translation of the given hypergraph description into a graph-based support,
and therefore provides no additional compression. On the other hand, the
inequality is strict when there are multiple equivalent hypergraph
parametrizations that lead to the same projected graph and functional rules. In
this case, the graph-based description eliminates the redundancy and achieves
better compression. The above therefore means that the marginal graph-based
encoding is, in general, at least as compressible as any individual
hypergraph-based choice.

This rather elementary consideration is important, since it shows that being
confined to graph-based representations is no impediment to achieving arbitrary
parsimony. Objectively, in this context, any inferential task amounts to
finding the best graph-based model that fits the data at hand. In some
situations, this description can be equivalently expressed with the aid of a
hypergraph, but such formulations are never strictly required---although either
representation may be more mathematically tractable, convenient, or
interpretable than the other in particular cases.

Finally, in most relevant empirical settings the available data $\D$ is finite,
and hence many different choices of $P(\D|\mathcal{M})$---involving also
representations $(\f,\W)$ and $(\g,\bl)$ that are not compatible with each
other---may offer competitive explanations for the same data, so that all these
aspects are ultimately relevant for model selection. Therefore, in practical
situations model selection may easily become a nuanced and computationally
demanding endeavor. However, it is important not to lose sight of the fact that
coupling multivariate interaction functions to the existence of hyperedges, as
done in Eq.~\ref{eq:hyperg} and in much of the HON literature, represents an a
significant structural and functional bias that may very easily not correspond
to the best modeling option. Indeed, as the example of Fig.~\ref{fig:graph-comp}
illustrates, a hypergraph model would not be able to accommodate multivariate
interactions if the interaction graph does not possess any cliques of the
corresponding order---an extreme modeling limitation that is not present in
graph-based descriptions. This consideration is important, for example, for the
problem of network reconstruction from time-series, where approaches based
strictly on hypergraphs may be blind to a wide class of structural and
functional patterns, and introduce statistically unjustifiable and hence
misleading artifacts in their outputs.

\section{Reducibility of multivariate interactions}\label{sec:reducibility}

As we have demonstrated, the recent literature on HONs conflates the existence
of multivariate interactions with hypergraphs, although they are perfectly
separable concepts. In addition, the concept of ``irreducibility'' of a
multivariate interaction is often evoked~\cite{battiston_networks_2020,bianconi_higher-order_2021,torres_why_2021,battiston_physics_2021,bick_what_2023,battiston_higher-order_2025,abiad_hypergraphs_2026}, as a means of justifying hypergraphs,
but never expanded in detail. Here we argue that this concept is far more
nuanced than the literature of HONs acknowledges.

We begin by observing that the dimension, or arity, of a function, i.e.\ the
number of arguments it takes, serves only a limited proxy for its complexity. In
particular, it is easy to show that every multivariate function can be expressed
as the sum of univariate terms, and, as a consequence, every multivariate
interaction in a network model can be expressed as a combination of pairwise
terms. The central fact that we invoke, discovered by
Cantor~\cite{cantor_ueber_1890}, is that the cardinality of the reals is
independent of dimension, i.e.\ $|\mathbb R^N| = |\mathbb R|$, which means that
there exist bijective functions of the form $\psi : \mathbb R^N \to \mathbb R$
that allow us to represent every multivariate function as an outer univariate
function applied on a projection of the arguments on the real line, i.e.
\begin{equation}
  f(\bm x) = h(\psi(\bm x)),
\end{equation}
with $h(y)$ being a univariate function. This fact can be used to develop
representations of multivariate functions in simpler terms. As an example, we
focus on the discrete bounded case $x_i \in \{0, \dots, A-1\}$, but our arguments
are extensible to more general scenarios. For such values, we can use the
following  bijection
\begin{equation}
  \psi(\bm x) = \sum_{i=1}^k\phi_i(x_i),
\end{equation}
based on a column-major ordering given by
\begin{equation}
  \phi_i(x_i) = A^{i-1}x_i,
\end{equation}
with the reverse mapping of $\phi(\bm x) = z$ given by
$x_i= \floor{(z \mod A^i) / A^{i - 1}}$. Based on this, we can write any
multivariate function as
\begin{equation}
  f(\bm x) = h\left(\textstyle\sum_i\phi_i(x_i)\right),
\end{equation}
where $h(z)$ is an appropriately defined univariate function. This means that
for a graph model, we could always write the interactions as
\begin{equation}\label{eq:decomp}
  f_i(x_i, \bm x _{\partial i}) = h_i\left(\phi_i(x_i) + \textstyle \sum_{j}W_{ij}\phi_j(x_j)\right),
\end{equation}
for some scalar function $h_i(z)$. Therefore, it is always possible to decompose
any multivariate function into sums of pairwise contributions---rendering moot
any formal distinction on this basis alone. We note that the bijection $\psi$ we
used is not unique, and there are other univariate functions $\phi$ that allow
the same representation of Eq.~\ref{eq:decomp}.

Superficially this may look like a simplification, but appearances can be
deceitful, since these scalar functions may be widely discontinuous and
arbitrarily complex, despite being one-dimensional. Although there are
alternative, more elaborate approaches such as Kolmogorov-Arnold
representations~\cite{andrey_kolmogorov_1957,arnold_representation_2009,braun_constructive_2009}
that achieve the same generality but with continuous univariate functions, in
the general case these functions will always be complex. Instead of arguing that
such univariate representations are universally useful, something we judge to be
unlikely, they are worth mentioning for the following reasons:
\begin{enumerate}
  \item The ``irreducibility'' of a multivariate function should not be taken
        for granted.
  \item For special cases, decompositions like Eq.~\ref{eq:decomp} may be simple and useful.
  \item The mere existence of a form like Eq.~\ref{eq:decomp} cannot be used to decide if an interaction is of ``low'' or ``high'' order, or ``reducible'' or ``irreducible.''
  \item Univariate representations can be useful for statistical inference, since only the individual univariate functions need to be parameterized.
\end{enumerate}
In fact, the last point is already being exploited in machine
learning~\cite{liu_kan_2024}, and can be potentially useful for network
reconstruction methods, in situations where the shape of the interaction
functions is not known beforehand.

In addition, there is another argument based on information theory which should
dispel the notion that multivariate functions, in particular those with a large
number of arguments, should not be expected to be simplified in some way. If we
keep ourselves to the same discrete case with $x_i \in X$, and functions that
map $X^k \to X$, with $|X|=A \in \mathbb N$ being the alphabet size, the number
of possible functions with $k$ arguments is given by $\Omega(A, k)= A^{A^k}$.
For a function not to admit a simpler representation, it needs to be
incompressible, which will be the case with overwhelming probability for a
function sampled uniformly at random from this set. To describe such a function,
we would need $\log_2 \Omega(A, k)= A^{k} \log _2A $ bits of information. Since
this number grows exponentially with the number of arguments, for $A=10$ we
would already need $\sim 10^5$ petabytes to describe an incompressible function
of $k=20$ arguments, $\sim 10^{14}$ petabytes for $k=30$, and $\sim 10^{84}$
petabytes for $k=100$. Since humanity never had and likely never will have
access to such large amounts of storage, it is safe to say that incompressible
functions are not only inaccessible for modeling, but are in fact quite useless
for this purpose. Consequently, the remaining set of functions that are
actually useful are compressible, which means their regularities are amenable to
modeling, and should not be considered as irreducible
monoliths~\cite{battiston_higher-order_2025}. In fact, this insight should hold
in general whenever higher-order representations are considered: the higher the
order of a representation, the more structured and hence decomposable it needs
to be, to be of any use for science as plausible descriptions of the world.

\section{Conclusion}\label{sec:conclusion}

We have provided conceptual, mathematical, and historical evidence refuting the
notion---prominent in the recent HON
literature~\cite{battiston_networks_2020,bianconi_higher-order_2021,torres_why_2021,battiston_physics_2021,bick_what_2023,battiston_higher-order_2025,abiad_hypergraphs_2026}---that
hypergraph models supersede graph-based representations for describing
multivariate interactions. Specifically, we showed that graphs do not merely
encode ``pairwise interactions.'' Rather, graphs specify \emph{with whom} a node
interacts, not \emph{how} those interactions are realized. Graph
parameterizations leave the interaction functions themselves unconstrained, and
the network science literature has long explored a wide range of such functions,
many of which require coordination and cooperation among multiple neighbors and
therefore cannot reasonably be classified as pairwise.

By the same logic, the notion that hypergraphs are more general does not hold up:
hypergraph parameterizations can only impose additional constraints on how
interactions are specified, and thus constitute strict special cases of more
general graph-based formulations. Consequently, graph formulations should not be
universally regarded as a ``reduction'' of a hypergraph, nor should a hypergraph
be viewed as an ``extension'' of a network; in both cases, the reverse is
generically true.

This observation has important implications for several aspects of network
modeling and analysis, particularly for network reconstruction from time-series
and other indirect
data~\cite{peixoto_network_2019,peixoto_network_2025,peel_statistical_2022}.
Rather than developing explicit but ill-defined ``generalizations'' to
hypergraphs, it suffices to relax the constraints on the latent node functions,
as this approach is already maximally general.

Furthermore, we demonstrated that the central phenomenological claims purported
to be uniquely attributable to hypergraph parameterizations often rest on
calculations that effectively disregard the hypergraph structure itself. As a
result, these analyses produce outcomes identical to those obtained from locally
tree-like graph models lacking cliques---let alone genuine hyperedges. The
reported behaviors therefore fall within previously established classes of
emergent phenomena, such as abrupt transitions in cooperative dynamics, rather
than constituting a genuinely new class.

Based on these observations, researchers should not proceed under the
misapprehension that hypergraphs are necessary, or even generically useful, to
study complex systems. More concretely, conflating multivariate interactions
with hypergraph formulations can introduce significant biases into theories and
methods that are developed under the tacit---but incorrect---assumption that
these concepts are inseparable.

Historically, network science has experienced waves of generalizations aimed at
developing more realistic theories. Notably, substantial effort has gone into
understanding networks that evolve in time~\cite{holme_temporal_2012} and
networks composed of multiple interaction
layers~\cite{de_domenico_mathematical_2013}. The current HON literature appears
to be modeled after these developments. However, it differs in a crucial
respect: the dynamic and multilayer nature of network systems is empirically
indisputable. It is not debatable whether networks change over time, or whether
multiple interaction modalities matter---for example, whether a disease spreads
via ground or air transportation has a profound effect on its dynamics.

The situation is fundamentally different for so-called ``higher-order
networks.'' Here, the motivation rests on overlooking the capacity of network
models to represent multibody interactions---and on the unjustified assumption
that hypergraph parameterizations expand the space of possible interactions,
when, in fact, they contract it. As we have argued, it is therefore unsurprising
that empirical evidence supporting the most extreme claims of generality and
ubiquity of hypergraph effects remains extremely sparse.

We emphasize that we are not arguing that hypergraph parameterizations are
uninteresting or without value. On the contrary, they have proven useful, for
example, in statistical physics~\cite{mezard_information_2009}, topological data
analysis~\cite{wasserman_topological_2018} and signal
processing~\cite{schaub_signal_2021}. Moreover, hypergraphs have historically
been understudied in network science, and renewed interest in them is well
justified on that basis alone. It is also reasonable to expect that hypergraphs
may offer more natural, interpretable, or parsimonious representations of
specific classes of systems. What we object to are unjustified claims about
their overall scope and necessity, and inaccurate characterizations of the
results obtained using them.

We stress also that the study of higher-order or multi-body interactions
\emph{proper}---such as the dependence of the population of one species on a
nonlinear combination of two or more others, the regulation of a gene depending
on a hierarchy of promoters or inhibitors, the modulation of a chemical reaction
between two reactants by a catalyst, or the dependence of a dynamical system on
its history~\cite{lambiotte_networks_2019}, and so on---is an extremely
important endeavor. This line of inquiry dates back to the origins of the field
and is not the target of our critique. Rather, our criticism is directed at a
persistent conflation in the recent literature that equates such multivariate
interactions with hypergraph models. As we have shown, these two notions are
entirely separable. Recognizing this distinction clarifies the modeling
landscape and opens the door to leveraging the complementary strengths of both
graph-based and hypergraph-based approaches for studying multivariate
interactions.

\bibliography{bib,mdd}

\appendix

\section{Equilibrium models, Hamiltonians, and factor graphs}\label{sec:hamiltonians}

The Hamiltonian $H(\bm x)$ of a system defines the joint distribution of $N$
variables, up to normalization, i.e.
\begin{equation}
  P(\bm x) = \frac{\ee^{-H(\bm x)}}{Z},
\end{equation}
with the partition function $Z=\sum_{\bm x}\ee^{-H(\bm x)}$ serving as a
normalization constant. A common parameterization of Hamiltonians is a
bipartite factor graph~\cite{mezard_information_2009}, defined as a set $F$ of
factors $a\in F$ that represent functions over a subset $\partial_a$ of the
variables, i.e.
\begin{equation}
  H(\bm x) = \sum_{a\in F}f_a(\bm x_{\partial_a}).
\end{equation}
Since bipartite graphs and hypergraphs are equivalent mathematical objects, this
can be seen as a hypergraph formulation. This formulation is indeed very
fruitful, as it can be used to develop message passing algorithms to efficiently
characterize properties of the distribution, which become exact when the factor
graph is a tree, and give excellent asymptotic approximations when it is locally
tree-like~\cite{mezard_information_2009}. Nevertheless, as before, such
formulations can be seen as a special case of the more general graph-based
framework given by
\begin{equation}\label{eq:ham}
  H(\bm x) = \sum_{i}f_i(x_i,\bm x_{\partial_i}).
\end{equation}
This is true because every factor graph formulation can be represented by
\begin{equation}
  f_i(x_i,\bm x_{\partial_i}) = \sum_{a\in F}\bm{1}_{i \in \partial_a}f_a(\bm x_{\partial a}),
\end{equation}
while the same decomposition is not possible in general in the reverse direction
for an arbitrary function $f_i$.

One could correctly argue that Eq.~\ref{eq:ham} is also an instance of a factor
graph, with one factor per node connecting it to its neighbors. The key point is
that this formulation is maximally general, relying solely on the graph’s
adjacency structure. Merging two factors incident on the same node would
effectively increase its adjacency, so this representation is already as general
as possible for a given adjacency pattern.

As a concrete illustration of the distinction we make above, we consider the
slight generalization of the hypergraph Ising model of
Ref.~\cite{robiglio_synergistic_2025}, with a Hamiltonian given by
\begin{equation}\label{eq:cham}
  H(\bm x) = -\sum_ih_ix_i - \sum_{\cc}\lambda_{\cc}\prod_{(i,j)\in \cc \times \cc}\delta_{x_i,x_j},
\end{equation}
where $\lambda_{\cc}$ is a coupling for hyperedge $\cc$, and $\bm$ $h_i$ is a
local field for variable $i$. This function can be re-written exactly as
\begin{equation}
  H(\bm x) = -\sum_ih_ix_i - \sum_i\sum_{\cc}\bm{1}_{i\in \cc}\frac{\lambda_{\cc}}{|\cc|}\prod_{j\in\cc}\delta_{x_i,x_j},
\end{equation}
where $|\cc|$ is the size of hyperedge $\cc$, which in turn is a special
case of the graph-based formulation of Eq.~\ref{eq:ham} that can be formulated
in many ways, without referring to hypergraphs. For example, we could write
\begin{equation}\label{eq:mham}
  H(\bm x) = -\sum_ih_ix_i - \sum_{i}\sum_l\prod_{j \in \partial_i(l)}W_{ij}^l\delta_{x_i,x_j},
\end{equation}
where $W_{ij}^l \in \mathbb R$ is the entry of a multilayer network, and
$\partial_i(l) = \{j \mid W_{ij}^l \neq 0\}$. From Eq.~\ref{eq:mham} we can recover
Eq.~\ref{eq:cham}, but not vice versa.

\end{document}